\documentclass[12pt]{iopart}
\usepackage{mathptmx}      % use Times fonts if available on your TeX system
\usepackage{flushend}
\usepackage[colorlinks,citecolor=blue,urlcolor=black,linkcolor=black]{hyperref}
\usepackage{rotating}
\usepackage[utf8]{inputenc}
\usepackage[english]{babel}

\usepackage[comma, authoryear]{natbib}
\usepackage{subcaption}
\usepackage{array}
\newcolumntype{L}[1]{>{\raggedright\let\newline\\\arraybackslash\hspace{0pt}}m{#1}}
\newcolumntype{C}[1]{>{\centering\let\newline\\\arraybackslash\hspace{0pt}}m{#1}}
\newcolumntype{R}[1]{>{\raggedleft\let\newline\\\arraybackslash\hspace{0pt}}m{#1}} 

\newcommand{\ops}{$\mbox{o-Ps}\rightarrow 3\gamma$ }

\usepackage{tikz}% no needs since pgfplots loads already it
\usetikzlibrary{spy}
\usetikzlibrary{positioning}

\begin{document}
\title{Feasibility study of the positronium imaging with the J-PET tomograph}

\author{
P.~Moskal$^1$, D.~Kisielewska$^1$,
C.~Curceanu$^2$, E.~Czerwi\'nski$^1$, 
K.~Dulski$^1$, A.~Gajos$^1$, M.~Gorgol$^3$,
%N.~Gupta-Sharma$^1$, 
B.~Hiesmayr$^4$, 
B.~Jasi\'nska$^3$, K. Kacprzak$^1$, \L.~Kap\l on$^1$, 
G.~Korcyl$^1$, P.~Kowalski$^5$, 
W.~Krzemie\'n$^6$, T.~Kozik$^1$, E.~Kubicz$^1$, 
M.~Mohammed$^{1,7}$,  Sz.~Nied\'zwiecki$^1$, M.~Pa\l ka$^1$,  M.~Pawlik-Niedźwiecka$^1$,
L.~Raczy\'nski$^5$, J. Raj$^1$, 
%K.~Rakoczy$^1$, Z.~Rudy$^1$, 
S.~Sharma$^1$, Shivani$^1$, R.~Y.~Shopa$^5$,
M.~Silarski$^1$, M.~Skurzok$^1$, E.~St\c{e}pie\'n$^1$,
W.~Wi\'slicki$^5$, B.~Zgardzińska$^3$
%M.~Zieli\'nski$^1$
\address{$^1$ Faculty of Physics, Astronomy and Applied Computer Science, Jagiellonian University, 30-348 Cracow, Poland}
\address{$^2$ INFN, Laboratori Nazionali di Frascati, 00044 Frascati, Italy.}
\address{$^3$ Institute of Physics, Maria Curie-Sk\l odowska University, 20-031 Lublin, Poland}
\address{$^4$ Faculty of Physics, University of Vienna, 1090 Vienna, Austria}
\address{$^5$ Department of Complex Systems, National Centre for Nuclear Research, 05-400 Otwock-\'Swierk, Poland}
\address{$^6$ High Energy Physics Division, National Centre for Nuclear Research, 05-400 Otwock-\'Swierk, Poland}
\address{$^7$ Department of Physics, College of Education for Pure Sciences, University of Mosul, Mosul, Iraq}
}            
\begin{abstract}
A detection system of the
conventional PET tomograph is set-up to record data from $e^+e^-$ annihilation into two
photons with energy of~511~keV, 
and it gives information on the density distribution
of a radiopharmaceutical in the body of the object.
In this paper we explore the possibility of performing the three gamma photons 
imaging based on ortho-positronium annihilation, as well as the possibility 
of positronium mean lifetime imaging with the \mbox{J-PET} tomograph constructed from plastic scintillators. 
For this purposes simulations of the ortho-positronium formation 
and its annihilation into three photons were performed taking into account 
distributions of photons' momenta as predicted by the theory of quantum electrodynamics and the response of the J-PET tomograph.
In order to test the proposed ortho-positronium lifetime image reconstruction method,
we concentrate on the decay of the ortho-positronium into three photons
and applications of radiopharmaceuticals labeled with isotopes emitting a prompt gamma.
The proposed method of imaging 
is based on the determination of hit-times and hit-positions of registered photons 
which  enables the reconstruction of the time and position of the annihilation point as well as the lifetime of the ortho-positronium on an event-by-event basis.
%The annihilation point and time of the $e^+e^- \to 3\gamma$ processes is reconstructed in an
%analytical way based on the measured times and positions of the three photons, and knowing that (due to the momentum conservation principle) all three photons
%momentum vectors and the annihilation point are lying in a single plane.
We have simulated the production of the positronium in point-like sources and in a cylindrical phantom composed of a set of different materials in which the ortho-positronium lifetime varied from 2.0~ns to 3.0~ns, as expected for ortho-positronium created in the human body.     
The presented reconstruction method for total-body \mbox{J-PET} like detector allows to 
achieve a mean lifetime resolution of $\sim$40~ps. Recent Positron Annihilation Lifetime Spectroscopy  measurements of cancerous and healthy uterine tissues
show that this sensitivity may allow to study the morphological changes~in cell structures.
\end{abstract}

\maketitle

\section{Introduction}
\label{sec::introduction}
%Positron Emission Tomography (PET) allows to determine spatial and sometimes also temporal distribution of
%concentrations of selected substances in the body. To this end, the patient is administered 
%pharmaceuticals marked with radioactive isotope emitting positrons 
%(e.g. $^{11}C$, $^{13}N$, $^{15}O$, $^{18}F$, etc.). The emitted positron %travels a very short distance in  tissues, where it
%annihilates with electron from a tissue component. 

Positron emitted inside the human body 
%(during $\beta^+$ decay of the radio-tracer) 
can either annihilate directly with one of the electrons of the examined organism or it creates
the metastable state of electron and positron called positronium~(Ps).
Positronium may be trapped inside free volumes between and within molecules of the examined patient. 
Currently, in the commercial PET technique, the phenomenon of positronium production
is neither recorded nor used for imaging.
%The detection system of the conventional PET
%tomograph is set-up to record data on the annihilation into two photons with energy of 511~keV,
%and it gives information on the density distribution of a radiopharmaceutical 
%in the body of the object~\citep{Karp,Conti,Slomka,Vandenberghe2016}.
%The parameter determining the degree of metabolic changes recorded by PET is the SUV index
%(Standardised Uptake Value), which expresses the value of the uptake of the radiopharmaceutical
%in a volume unit (voxel) of the organism in relation to the average value of the uptake of the
%radiopharmaceutical throughout the body. The higher the SUV the greater is the probability of
%occurrence of cells with altered metabolism in a given tissue region. 

Imaging of the properties of positronium inside the body may deliver new diagnostic information. 
The average lifetime of positronium depends on the size of the free volumes between 
atoms~\citep{Schrader:1998,Coleman:2000,Tao:1972,Eldrup:1981} 
and there are indications~\citep{JasinskaActaB:2017,JasinskaActaA:2017,Pietrzak,Liu2007} 
that it is correlated with the stage of the development of metabolic disorders of the human tissues.
Therefore, an image of the average lifetime and production probability of the positronium formed
inside the human body during the routine PET imaging~\citep{patentMoskal},
as well as the fraction of its annihilations into three-photons~\citep{Kacperski2005,Jasinska-Moskal2017}
may deliver information complementary to the standardized uptake value (SUV) and useful for the diagnosis.  
\begin{figure}[h]
    \centering
    \includegraphics[width=0.6\textwidth]{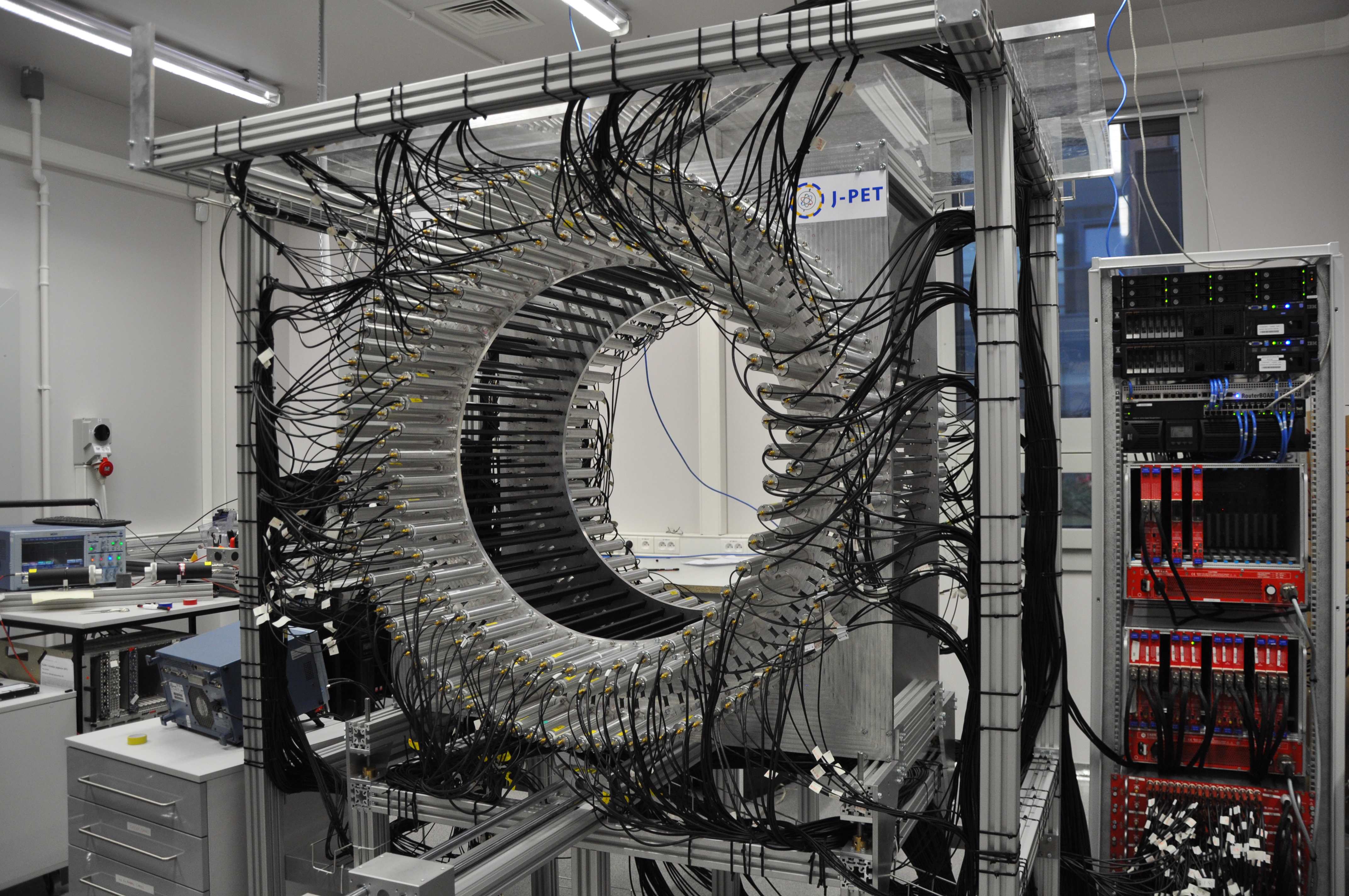}
    \caption{Photo of the prototype of the Jagiellonian Positron Emission Tomograph 
         (\mbox{J-PET}) scanner. The J-PET scanner is built from strips of 
         plastic scintillator
         with dimensions of $7  \times 19 \times 500 \mbox{ mm}^3$,
         forming cylindrical layers~\citep{Szymon-Acta,Moskal:2014rja,Moskal:2014sra}
         with the inner diameter of 85~cm and the axial field-of-view (AFOV) of 50~cm.
         Light signals from each strip are converted
         to electrical signals by Hamamatsu R9800 vacuum tube photomultipliers placed 
         at opposite ends of each strip.  Note that the present J-PET prototype consists of 192
         scintillator strips only, with  empty spaces between the scintillators.
     \label{fig::JPETphoto}
     }
\end{figure}
In this article we explore a possibility of positronium average lifetime imaging with the \mbox{J-PET} 
tomograph constructed from plastic 
scintillators~\citep{Moskal:2016ztv,Szymon-Acta}. 
We concentrate on the decay of the ortho-positronium (o-Ps) into three photons
and applications of radiopharmaceuticals emitting prompt gamma. 

In more than 40\% cases, the electron-positron annihilation proceeds in the tissue 
via creation of positronium~\citep{Harpen2004,JasinskaActaB:2017}
and even in water the ortho-positronium is formed with the probability of about 25\%~\citep{Stepanov2011}.
Ortho-positronium decays in vacuum into three photons, however, in the human body, 
due to the large probability of the pick-off processes~\citep{Garvin1571},
the three-photon events (both from ortho-positronium decays and direct 
$e^+e^- \rightarrow 3\gamma$ processes) constitute altogether only about 0.5\% of 
electron-positron annihilations~\citep{JasinskaActaB:2017,Jasinska-Moskal2017}.
However, such a relatively low fraction of three-photon annihilations occurring in the
human tissues may be compensated by increased detection efficiency of the PET detector and 
by improved spatial resolution of the single-event annihilation point reconstruction~\citep{Gajos:2016nfg,Kacperski2005}.
At present, PET tomographs record and use for image reconstruction
less than 1\% of annihilation events occurring 
in the body~\citep{Cherry2017, Cherry2018}, 
%thus leaving room for significant improvement of three-photon statistics by the increase of sensitivity, 
%without any need to extend the duration of examinations or to increase the dose of radiopharmaceuticals.
In this context it is important to emphasize that there is an ongoing development of novel modalities 
for the whole and total-body PET imaging which shall be characterized by about 
40~times higher sensitivity~\citep{Viswanth2017,Zhang2017,Cherry2017,Cherry2018}
% Viswanth2017 -bibitem wygląda ok
with respect to the presently available PET systems~\citep{Slomka,Vandenberghe2016}. Moreover, additional improvement in the time resolution to 100~ps, combined with the total-body PET,
may considerably increase the sensitivity even by a factor of 200~\citep{Cherry2018}. 
Therefore, an efficient positronium imaging may become feasible in the nearest future. 

The reconstruction of positronium lifetime requires determination of times of its creation and annihilation.
These can be achieved when applying radiopharmaceuticals labeled with isotopes as e.g. Scandium-44 which after emission of the positron changes into a daughter nucleus in an excited state~\citep{Walczak2015,Szkliniarz2015}.
The daughter nucleus subsequently de-excites through emission of one or several gamma quanta (often referred to as prompt gamma).
Thus, as a result of the decay both:  $e^+$ and $\gamma$  are emitted e.g. 
via the following process: 
$^{44} \mbox{Sc} \rightarrow ^{44}\mbox{Ca}^{*} \ e^+ \ \nu_e \rightarrow ^{44}\mbox{Ca} \ \gamma \ e^+ \ \nu_e$,
where, $\nu_e$ denotes a neutrino whose interaction probability with the patient and detector is negligible.

\begin{figure}[tp]
    \centering
    \includegraphics[width=0.55\textwidth]{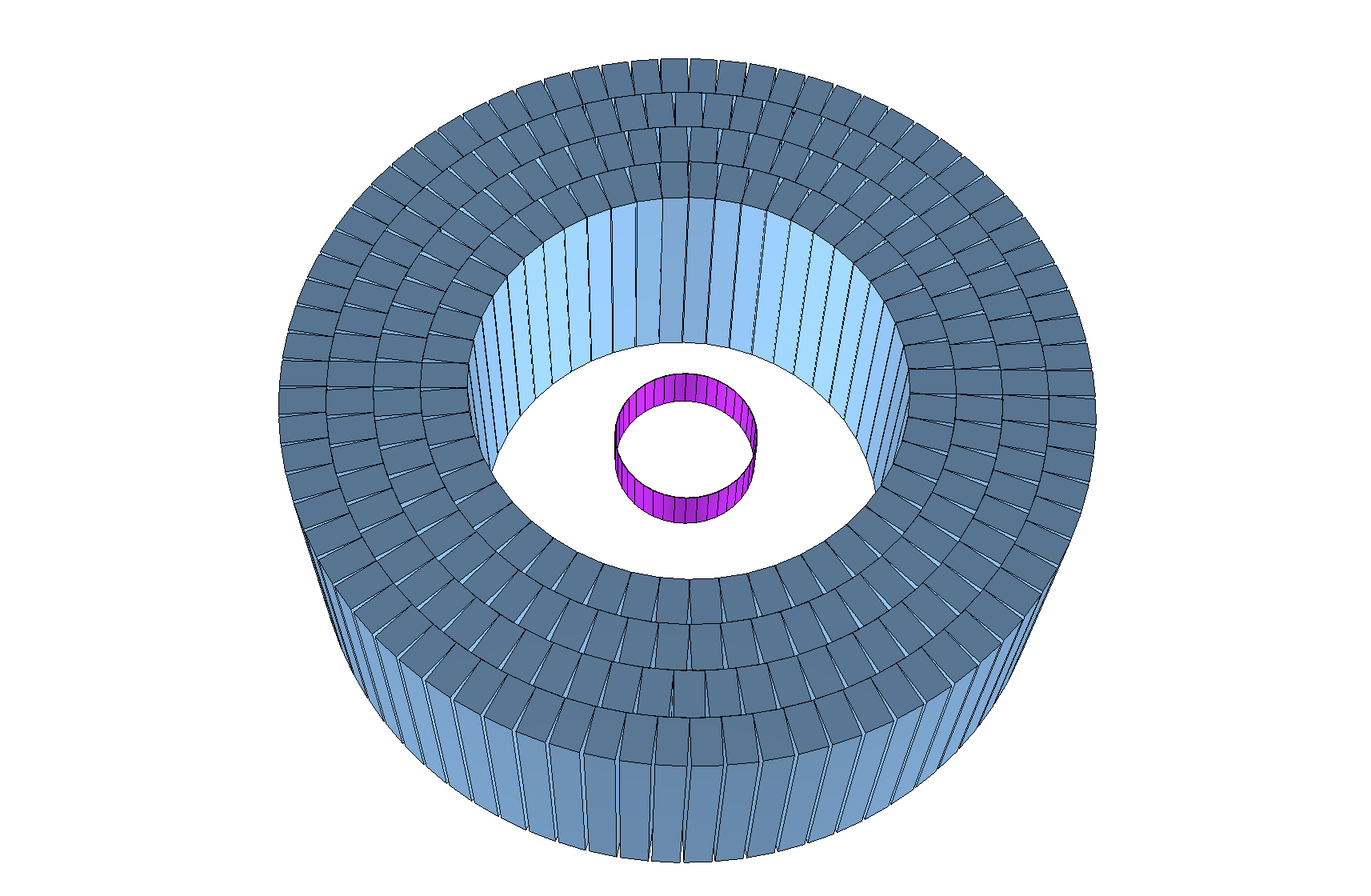}
    \includegraphics[width=0.35\textwidth]{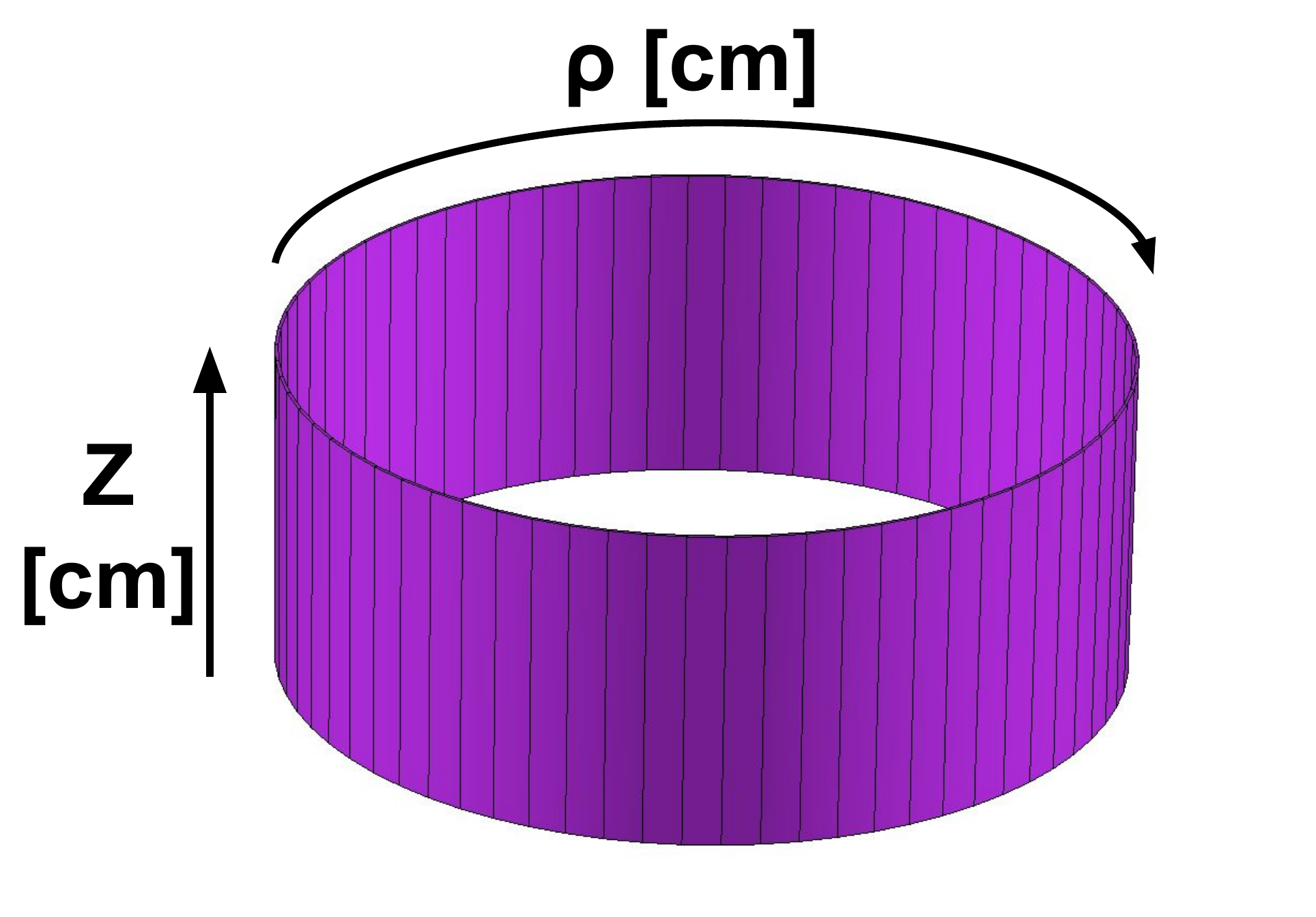}
    \caption{Left: visualization of simulated cylindrical shape detector. 
    Detector consists of four layers and each of them is filled with 
    scintillators. In this article the simulations were performed assuming scintillators 
    dimensions of $7 \times 19  \times 500 \mbox{ mm}^3$ 
    and the inner layer radius of 43~cm.
    However, for clarity the elements are shown not to scale.
    In each layer only a selected number of scintillators
    with increased x-y dimensions is shown.
    %The simulated distances between the center of the detector and the  consecutive 
    %layers and the number of the scintillators in each layer are listed in 
    %Table~\ref{tab::detector}.
    Inside the detector a cylindrically shaped phantom is placed. 
    Right: visualization of simulated phantom 
    ($R_{inner}=10.0$~cm, $R_{outer}=10.1$~cm, $z=10.0$~cm) with  coordinate system as described in the text. The phantom is composed of various materials in which the
    lifetime of ortho-positronium atoms varies from 2200~ps to 2300~ps as expected 
    for the positronium in the human body. A comprehensive discussion of the known results 
    on o-Ps lifetime can be found in section 2.
    \label{fig:oPs_detector}
    }
\end{figure}

Recently, applications of radiopharmaceuticals labeled with metallic $\beta^+$
emitters as radioisotopes of scandium ($^{43}$Sc, $^{44}$Sc) have rapidly expanded, 
particularly for oncologic imaging purposes~\citep{Hofman2012,Krajewski2013,Huclier2014,Walczak2015}.
For example the $^{44}$Sc labeled DOTATATE due to its high binding affinity for the somatostatin
receptors can enhance significantly diagnostics quality of neuroendocrine cancers~\citep{Sing2017}.

\begin{figure}[tp]
    \centering
    \includegraphics[width=0.9\textwidth]{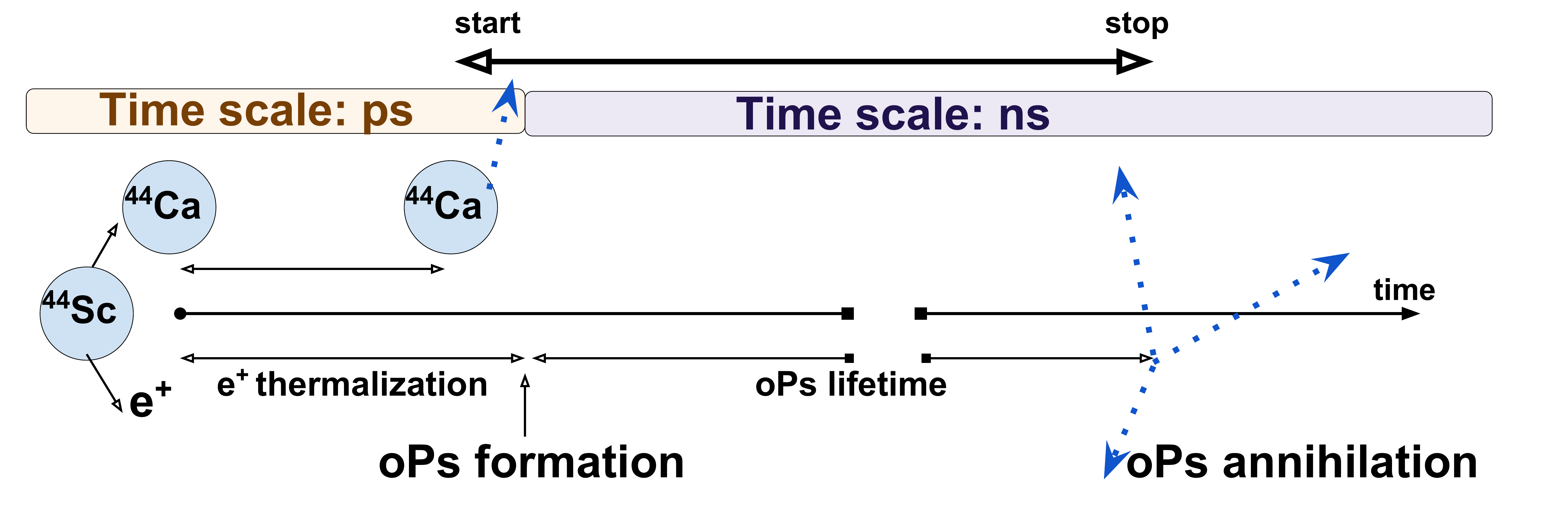}
    \caption{ Scheme of scandium $\beta^+$ decay that leads to formation of ortho-positronium state.
       The  $^{44}$Sc isotope emits a positron which thermalizes in the body and forms an ortho-positronium bound state. 
    Independently, the remaining $^{44}$Ca* nucleus deexcites after a few picoseconds~\citep{Nucleide:LARA} 
    emitting a  prompt gamma  with  characteristic energy.
    Both the time scale of thermalization~\citep{Schrader:1998} 
    and $^{44}$Ca excited state lifetime are
    in the order of picoseconds, while the lifetime of created ortho-positronium is in the order of nanoseconds.
    This allows to treat the de-excitation photon emission time as the time when the
    ortho-positronium was formed. The o-Ps annihilation time is reconstructed
    based on gamma photons hit-times and hit-positions in the detector.  
    \label{fig:Sc_decay}
    }
\end{figure}

In the current PET imaging the prompt gamma constitutes a source of background since when reacting via Compton effect
it may give a signal in the detector which can be misclassified as signal 
from a 511~keV photon from the electron-positron annihilation. 
%(see Figure~\ref{fig:oPs_detector}, bottom panel). 
Analogously, in typical clinical PET, positron-electron annihilation to three
photons ($e^+ e^- \rightarrow 3\gamma$) constitutes a source of unwanted
background which is discarded together with the multi-photon events due to the
random coincidences and scatterings in the detectors. 

However, the triple coincidences originating from $e^+ e^- \rightarrow 3\gamma$
processes or occurring when utilizing the $\beta^+\gamma$ emitters may be useful
for imaging. 
Since more than a decade, there is also an ongoing research aiming at improvement of
spatial resolution of the annihilation point reconstruction when using $\beta^+\gamma$ 
emitters by combining standard PET tomography technique with the various types of Compton cameras
for the registration of the prompt gamma~\citep{Lang2012,Lang2014,Thirolf,Gringon2007,Donnard2012,Oger:2012}.
Also a fraction of detector scatter coincidences may be applied to
improve the image by recovering true Lines of Response (LORs) from triple 
coincidences~\citep{recoveringTriple,simulateTriple}. 

Image reconstruction based on the three-photon annihilation was also proposed more than decade ago 
by~\citep{Kacperski2004,Kacperski2005,Kacperski2007}.
The proposed method enables a reconstruction a place of the $e^+ e^- \rightarrow 3\gamma$ annihilation based 
on the measurement of energies of three photons and positions of their interactions in the detector material.
The application of energy and momentum conservation laws allows to reconstruct the annihilation point on an event-by-event basis.
However, the precise measurement of photon energies requires applications of 
high-resolution semiconductor detectors 
(such as e.g. CdZnTe~\citep{Kacperski2005,Kacperski2007}) 
rendering the implementation of the full scale PET scanner relatively expensive.

\begin{figure}[b]
	\centering
    \includegraphics[width=0.7\textwidth]{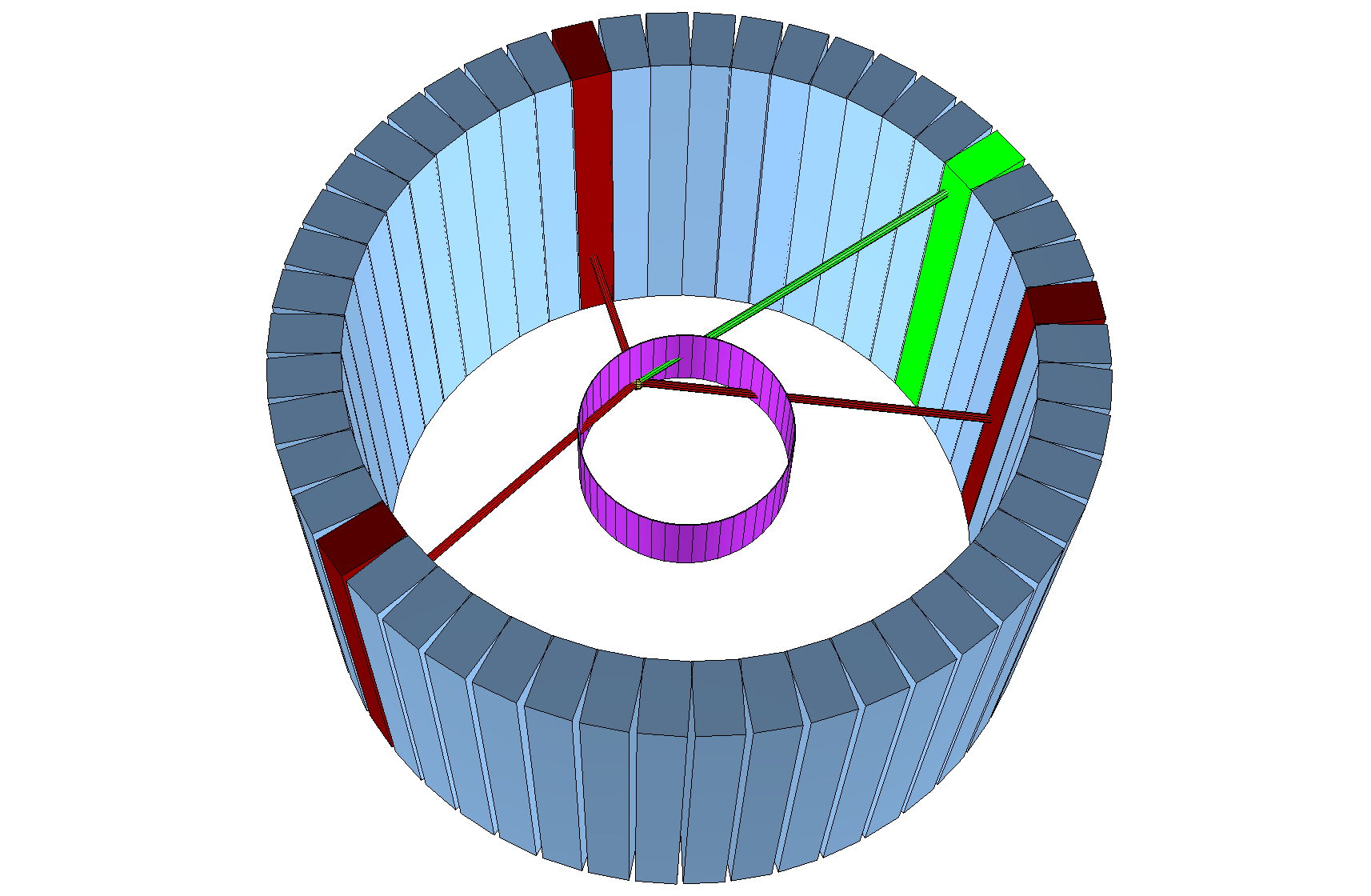} 
    \caption{A scheme of a simulated detector and  phantom located inside it with
    an exemplary $e^+ e^- \rightarrow 3\gamma$ annihilation event.
    For clarity only a single layer of detector is shown with decreased number 
    of scintillators and increased x-y dimensions.
    Red lines represent the photons originating from
    ortho-positronium annihilation, while green line shows prompt gamma
    originating from $\beta^+$ emitter de-excitation, 
    e.g. from $^{44}Sc \ \to \ \  ^{44}Ca^*\ e^+\ \nu \ \ \to \ \ \  ^{44}Ca \ \gamma \ e^+ \ \nu$   
decay chain.
    Photon recording strips are marked with corresponding colors.
    \label{fig::detector_with_ops}
    }
\end{figure}

In this article we present a cost-effective method enabling reconstruction
of the density distribution of the $e^+ e^- \rightarrow 3\gamma$ annihilation
points, as well as reconstruction of the image of ortho-positronium properties,
such as its average lifetime and production probability.
The proposed imaging of ortho-positronium properties produced in the living 
organism is based on the time of signals registered with the plastic 
scintillators~\citep{patentMoskal}. 
Thus, the precision of the proposed method relies predominantly
on the time resolution of the tomograph
since the measurement of energy losses of registered photons is used only to disentangle signals between prompt gamma and annihilation photons. 
The measured times are used to reconstruct the hit-times and hit-positions of registered 
photons which  enables the reconstruction 
of the time and position of the annihilation point. A prototype of the tomograph enabling
application of this method has been built (Fig.~\ref{fig::JPETphoto})~\citep{Szymon-Acta,Dulski2018}
and validated for classical PET imaging~\citep{Monika-Acta}.

Here we explore the possibility of performing positronium imaging  using dedicated Monte-Carlo 
simulation procedures which are described in detail in references~\citep{Kaminska:2016fsn,Gajos:2016nfg}. 
The simulations are carried out assuming an ideal case, that the tomograph is built
from four concentric layers of axially arranged plastics scintillator strips adjacent 
to each other as indicated in Fig.~\ref{fig:oPs_detector}, left panel. 

In order to reconstruct a mean lifetime of ortho-positronium atoms in each voxel 
of the patient it is necessary 
%to identify events with ortho-positronium production
%and for each such event 
to determine the times of the ortho-positronium creation and decay (see Figure~\ref{fig:Sc_decay}).
The scheme of the detector with recorded \ops annihilation is shown in Figure~\ref{fig::detector_with_ops}.
The time of creation is determined using the prompt gamma e.g. from $^{44}Sc$, since nuclear deexcitation is much faster (few picosends) than positronium lifetime.

The annihilation point and time of $e^+ e^- \rightarrow 3\gamma$ processes is
reconstructed in an analytical way based on the measured times and positions 
of the three photons ($X_i$,$Y_i$,$Z_i$,$T_i$), $i$~=~1,2,3 and knowing that
(due to the momentum conservation principle) all three photons momentum vectors 
and the annihilation point ($x,y,z$) are lying in a single plane,
referred to as the decay plane~\citep{Gajos:2016nfg}.
Co-planarity of the $3\gamma$  reduces the reconstruction to a two dimensional problem of three equations:
($T_i - t$)$^2~c^2$ = ($X'_i-x'$)$^2$ + ($Y'_i-y'$)$^2$, 
where $X'_i$,$Y'_i$,$T_i$ (for i=1,2,3) denote hit-coordinates and hit-times 
in the decay plane reference system, $x'$, $y'$, $t$ denote the annihilation 
point's coordinates and time in this frame, and $c$ stands for the speed of light. 
Solving this system of three equations defined for
i~=~1,2,3 yields the time and  location point of the  $e^+ e^- \rightarrow 3\gamma$ annihilation.

\begin{table}[hb]
    \caption{Selected physical characteristics of the exemplary beta-plus isotopes 
        use for both: PET imaging and  
        positron annihilation lifetime spectroscopy (PALS)
        investigations.
       % For isotopes that decay into 
       %  excited states the properties of emitted gamma quanta are denoted.
         Data were adapted from~\cite{nncd}.}
    \label{tab::isotopes}
        %\begin{tabular*}{\textwidth}{@{\extracolsep{\fill}}|c|c|c|c|c|@{}} \hline
         \begin{tabular}{|c|c|c|c|c|} \hline
            Isotope & Half-life & $\beta^+$ decay & $E_{\gamma}$ [MeV] &
             Excited nuclei lifetime \\ \hline
            $^{44}$Sc & 4.0 h & $^{44}\mbox{Sc} \to ^{44}\mbox{Ca} + e^+ + \nu_e +\gamma$ &
            1.16  & 2.61 ps \\ \hline 
            $^{14}$O & 70.6 s & $^{14}\mbox{O} \to  ^{14}\mbox{N}+ e^+ + \nu_e + \gamma $ 
            & 2.31   & 67.8 fs  \\ \hline
            %$^{22}$Na & 2.6 [year] & $^{22} \mbox{Na} \to ^{22}\mbox{Ne} + e^+ + \nu_e +\gamma$
            %& 1.27 & 0.546 &  3.63~[ps] \\ \hline
            %$^{18}$F &  1.8 [h] & $^{18}\mbox{F} \to  ^{18}\mbox{O} + e^+ + \nu_e$
            %& - & 0.634 & - \\ \hline
        \end{tabular}
    %}
\end{table}

The registration of the prompt gamma, in addition to the annihilation photons, permits to measure the differences between the time of formation and the time of 
annihilation of positronium on the event-by-event basis~\citep{patentMoskal,Gajos:2016nfg,Kaminska:2016fsn}.
Reconstructed time of the prompt gamma emission is used
as the formation time of the positronium.
This approximation leads to  negligible uncertainties, since the time 
between the emission of the positron and the formation of the positronium, 
as well as the average lifetime of the excited nuclei relevant for PET examinations 
(e.g. $^{44}Ca^*$), are both in the order of a few picoseconds (see Table~\ref{tab::isotopes}).

\begin{figure}[tp]
    \centering
     \includegraphics[width=0.49\textwidth]{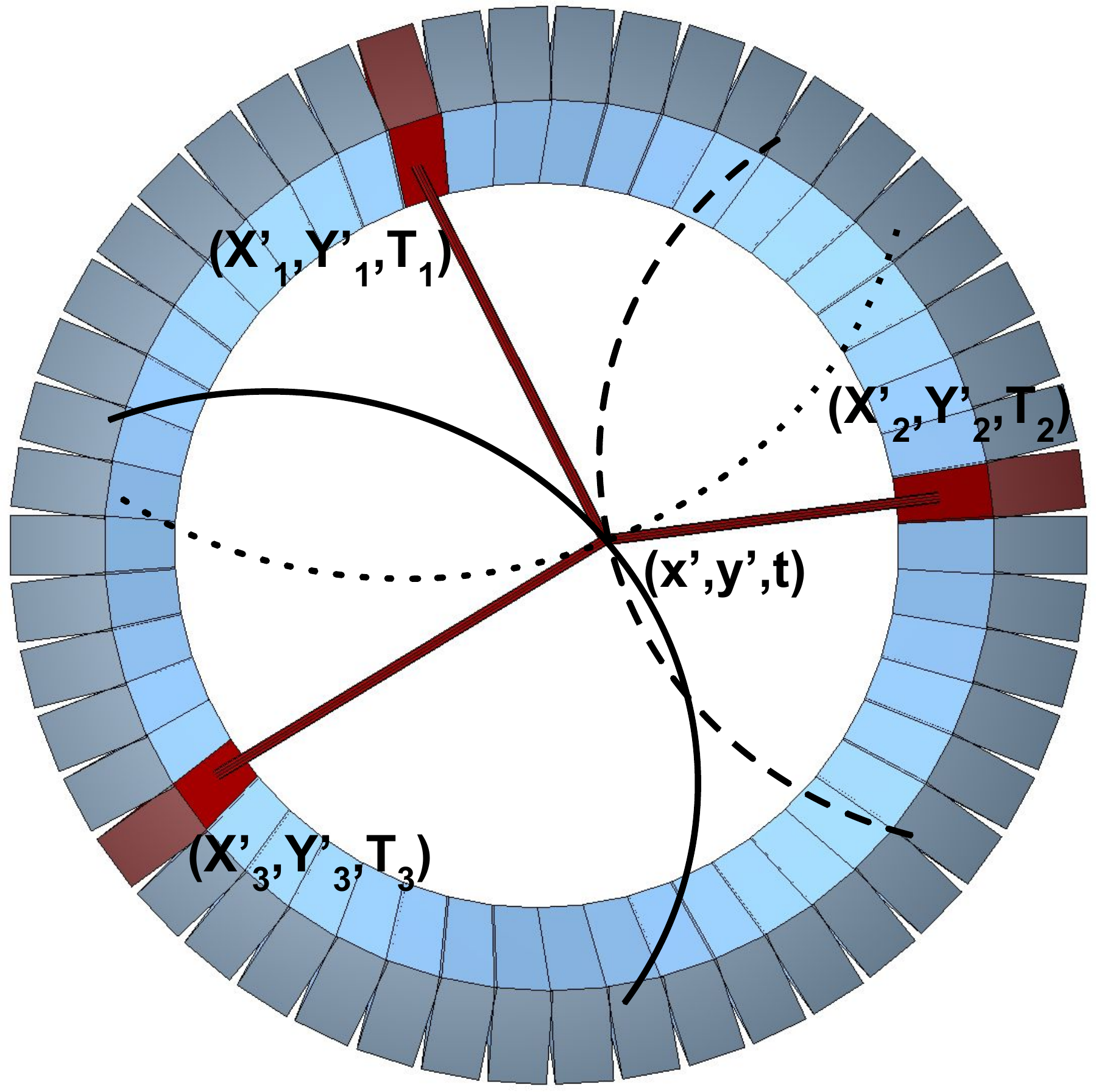}
     \caption{A scheme of the detector showing \ops annihilation.
    For clarity only a single layer with registered hits is shown.
    Red lines represent the gamma photons from ortho-positronium annihilation. 
     %The Plane of Response (in blue) is drawn for the annihilation photons.
    %Along the Plane of Response the 
     Figure shows a single layer of the
 detector in 3D view with cross-section of the scintillators shown in grey and
 their inner side in blue.
    The trilateration method is used to determine the annihilation 
    position and time $(x^{\prime},y^{\prime},t)$ along the 
    annihilation plane.
    For each recorded photon a circle, which is a set of possible photon origin points,
     centered in the hit-position 
     and
    parameterised with the unknown o-Ps annihilation time is considered. The intersection of the three circles corresponds
    to the \ops annihilation point.
    \label{fig:gps_reconstruction}
    }
\end{figure}

Finally, it is worth to note that 
if the time resolution would be improved in the future down to 10~ps,
as advocated by~\cite{Lecoq:2017}, then  $2\gamma$  pick-off annihilations 
may be used to determine the annihilation time and position for each event,
which  in coincidence with the registration of prompt gamma  would allow to
determine lifetime spectra for each voxel and, hence, to reconstruct the 
ortho-positronium lifetime based on $\mbox{o-Ps} \to 2\gamma$ pick-off 
annihilations. Such ortho-positronium pick-off annihilations for the human body (with the o-Ps lifetime of 
about $\tau_{tissue}\sim$2~ns) are about 70 times ($\frac{\tau_{vacuum}}{\tau_{tissue}} - 1$) more frequent than \ops events.   Therefore, the proposed method of positronium imaging shall become practical with the advent of total-body PET and improved time resolutions.

The article is organized as follows: 
 In section \ref{sec::positronium_in_body} 
 the positronium properties in vacuum and in the human body are briefly reviewed based on the performed experiments up to date. 
This section concludes with the estimation of the range and values of 
 lifetimes of  ortho-positronia produced in the various tissues of the human body 
($\sim$1.8~ns -- $\sim$2.5~ns), as well as in the expected differences in the ortho-positronium lifetime between the healthy 
and cancerous tissues ($\sim$50~ps~--~ $\sim$200~ps). 
The section~\ref{sec::jpet_description} contains a general description of the J-PET detector 
in order to illustrate the possibility of applying the introduced method
for the ortho-positronium lifetime imaging simultaneously to the regular PET scan. 
This section defines also the properties of the detection system for which the feasibility study of the positronium imaging will be performed in Section~\ref{sec::MC}. 
The description of the Monte Carlo simulation given in section~\ref{sec::MC} comprises: 
(i)~description of the four-layer detector built from 
axially arranged plastic strips adjacent to each other,
(ii)~three-photon  annihilation as predicted by the theory of quantum electrodynamics,
(iii)~gamma photon interaction within detector build 
from plastic scintillators 
%via Compton effect 
as a function of incident gamma photon energy,
(iv)~resolution for reconstructing position, time and energy loss of gamma photons in the detector, 
(v)~reconstruction of annihilation position based on the 
reconstructed gamma photon hit information. 
Next, Section~\ref{sec::results} includes details on positronium 
mean lifetime image reconstruction, 
%are presented. The article is completed with 
discussion about the achievable mean-lifetime 
resolution and positronium imaging sensitivity.

\section{Positronium in the  human body}
\label{sec::positronium_in_body}

Similarly to the hydrogen atoms, positronium 
may be formed in a singlet or triplet  spin states, referred to as para-positronium and ortho-positronium, respectively.
Due to the charge conjugation symmetry conservation, 
the para-positronium annihilates with emission of an even number of photons and 
ortho-positronium annihilation is subject to emission of an odd 
number of photons. Predominantly these are two and three photons for 
para- and ortho-positronium atoms, respectively.
The decay time of formed positronium depends on its quantum state: 
the observed lifetime for para-positronium localized in vacuum amounts to $\sim 125 \mbox{ ps}$,
while, due to the smallness of the fine structure constant and the phase-space volume available, in the decays of the
ortho-positronium its lifetime in vacuum is three orders of magnitude greater ($\tau_{vac.}$ = 142~ns)~\citep{AlRamadhan:1994,Vallery:2003,Jinnouchi:2003}.
When formed inside matter, positronium may be trapped inside volumes of lower electron density. However, the positronium trapped in a free volume is susceptible 
to the pick-off processes~\citep{Garwin:1953}, in which positron from the positronium
may annihilate with one of the surroundings electrons.
The smaller are the free voids between the atoms the larger is the probability 
of the pick-off process and, hence, the smaller is the mean lifetime of the trapped positronium. 
This effect is significant in case of ortho-positronium for which the mean lifetime
may change in the range from about 142~ns to even below 1~ns.
The correlation is described in the framework of well tested models which relate quantitatively 
the mean ortho-positronium lifetime with the size of the free void in the range from 0.2~nm 
up to 100~nm~\citep{Tao:1972,Eldrup:1981,Goworek:1997,Goworek:1998}.
In addition it is well known from the Positron Annihilation Lifetime Spectroscopy (PALS)
studies that the intensity of the 
production of ortho-positronium atoms is strongly correlated with the free voids 
size distributions and their concentration~\citep{Goworek:1978,Kobayashi:1989,Schrader:1998,Coleman:2000,Jasinska:2003,Jasinska:2003a}.

In the human body, positronium atoms can be created and trapped both, in the dense tissue and in the bio-fluids - importantly, including water. 
%Importantly for the proposed studies the positronium atoms may be created even in the water.
In  purified water the o-Ps lifetime amounts to 1.8~ns and its production probability 
is equal to about 25$\%$~\citep{Stepanov}.
There are few works demonstrating the application of PALS technique to model and living biological systems
reporting differences  between normal and cancerous cells and changes of the o-Ps lifetime during dynamical processes undergoing in cells
\citep{Jean1977,Liu2007,Jean2006,Jean2007,Liu2008,Zaynab2012,Axpe2014,Kubicz:2015}.
Recently, significant differences in PALS parameters between normal and diseased tissues
were also observed in samples of uterine leiomyomatis and normal muscles tissues 
taken from women-patients after surgery of uterus tumors~\citep{JasinskaActaB:2017,JasinskaActaA:2017}. 
%For all six studied patients it was found that the values of the mean ortho-positronium 
%lifetimes are about ~2~ns, and that they are larger for the tumorous tissues 
%than for the healthy ones by about 50~ps. At the same time, the intensity of positronium
%production was found to be smaller for altered tissues than for the normal ones~\citep{JasinskaActaB:2017}.
%The above discussed results indicate
The results of above quoted research indicate 
that in all investigations performed to date there is
a difference in PALS parameters determined for healthy and cancerous tissues. Thus, as suggested in references~\citep{patentMoskal,Jasinska-Moskal2017}, 
measurements of properties of ortho-positronium atoms 
(such as lifetime and production probability, 
or $3\gamma/2\gamma$ rate ratio)
which are formed inside the
human body during a routine PET imaging may deliver information useful for the diagnostics.
In this article as a first step on a way towards realisation of such measurements, 
we present a feasibility study of the reconstruction method of the image of ortho-positronium lifetime proposed in reference~\cite{patentMoskal}.  

The mean ortho-positronium lifetimes in the tissues varies from about 1.8~ns (as in pure water) 
to about 2.5~ns (or even up to about 4~ns as measured for the human skin with low energy positron beam~\citep{Chen:2012}).
Whereas, the mean ortho-positronium lifetime differences for healthy and cancerous
tissues are in the range of about 50~ps to about 200~ps~\citep{JasinskaActaB:2017,Pietrzak,Liu2007}.

Taking into account the above information, the feasibility study of the positronium lifetime
imaging will be performed assuming the formations of ortho-positronium atoms in a
set of point-like sources and in a 
virtual phantom formed from tissues in which ortho-positronium lifetime varies from 2.0~ns to~3.0~ns. The phantom will be inserted into an idealized 
version of the J-PET tomograph consisting of four cylindrical layers.
In the simulations the properties of the detection system, which are briefly described in the following section, will be taken into account.

\section{Materials and methods}
\subsection{Principles of operation of the J-PET detector made of plastic scintillators}
\label{sec::jpet_description}

\begin{figure}[htp]
    \centering
    \includegraphics[width=0.45\textwidth]{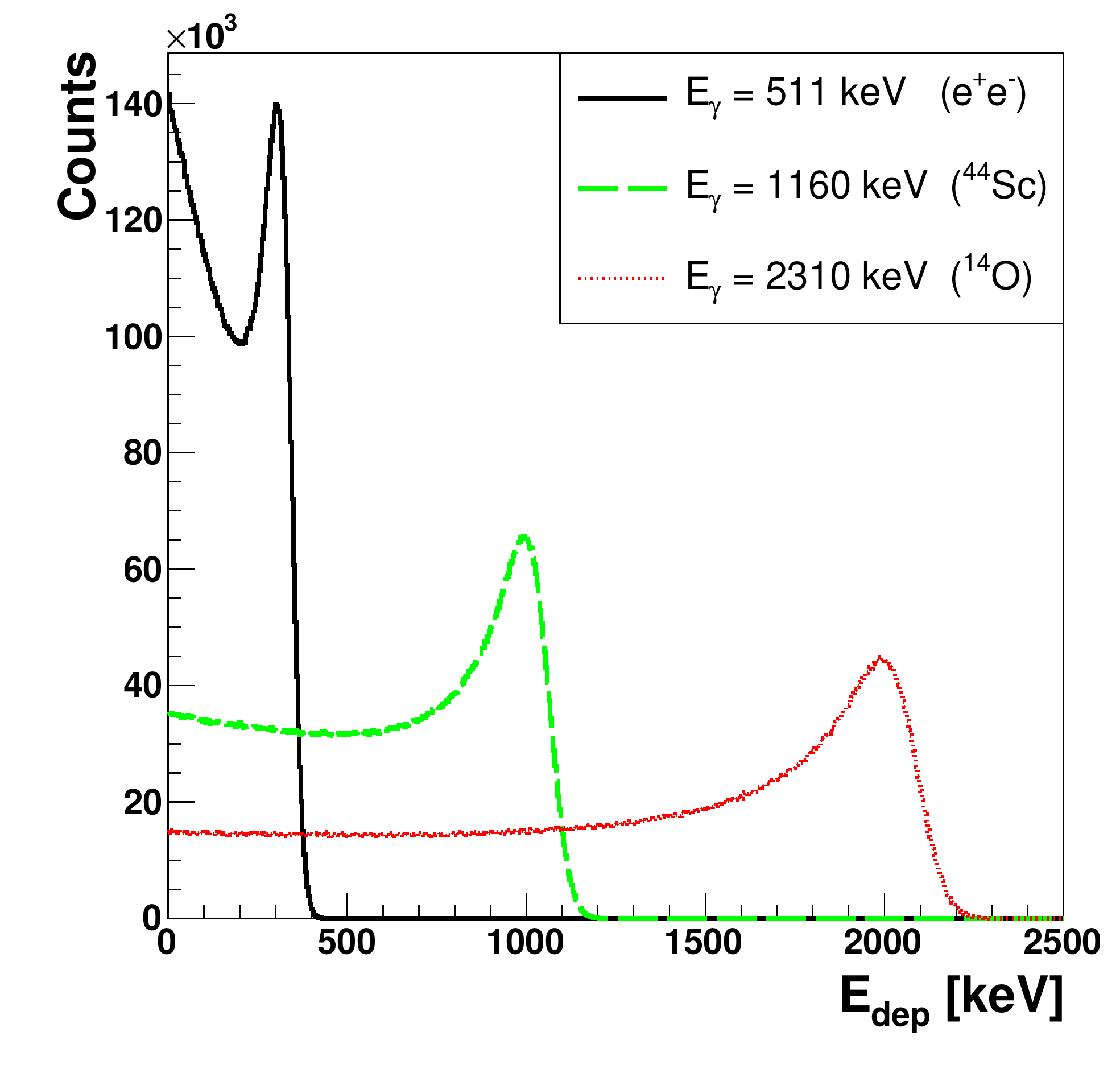}
    \caption{ 
        %Left: 
        Simulated energy loss spectra for annihilation photons (511 keV)
        and prompt gamma from the decay of isotopes indicated in the legend. The shown distributions include the energy resolution of the J-PET detector~\citep{Moskal:2014sra}. 
        %Right: 
        %Comparison between simulated and measured energy loss distributions for the 511~keV gamma
        %quanta. The simulated spectrum was normalized to the experimental one in amplitude. The lower range of
        %the experimental spectrum is cut by the threshold set and the slight enhancement of experimental spectrum over the simulated one around 400~keV is due to the multiple gamma photon scattering in the  scintillator. 
        \label{fig::spectra}
    }  
\end{figure}

The J-PET tomograph (shown in Fig.~\ref{fig::JPETphoto}) is a cylindrically shaped PET detector made of plastic scintillators. 
The light signals are converted to electric pulses by photomultipliers
connected optically at the two ends of a scintillator.
%as indicated in case of two modules in Fig.~\ref{fig:two_strips}.
In plastic scintillators, which are composed of elements of low atomic number, incident
photon interacts mainly through the Compton scattering leading to a continuous  energy loss distribution. Fig.~\ref{fig::spectra} presents an example of the
energy loss spectra for photons from $e^+e^- \to 2\gamma$ annihilation and for prompt gamma 
from $^{44}Sc$ and $^{14}O$. 
%The right panel of the figure indicates that there is a satisfactory agreement between experimental and simulated distributions~\citep{Moskal:2014sra}. 

In the J-PET tomograph the electric signals from photomultipliers are 
processed by the front-end electronics based on Field Programmable Gate Arrays (FPGA)~\citep{Palka:2017wms}, collected by means of the triggerless data acquisition system~\citep{Korcyl:2016pmt,Korcyl:2018}, and analysed using a dedicated 
analysis software framework~\citep{Krzemien:2015}.

%\begin{figure}[htp]
 %   \centering
%    \includegraphics[width=0.4\textwidth]{fig/figure6.eps}
%    \caption{Scheme of two plastic scintillator strips used in the J-PET %detector, where green cylinders
%    at the ends denote photomultipliers.
%  Yellow dot denotes the annihilation point and the thick black arrows %indicate annihilation photons.
%        Photons' interaction time ($t_1, t_2$, violet color) and positions %along the strips ($\Delta z_1$, $\Delta z_2$)  
%        are determined based on registered  times of scintillation light %arrival at the ends
%         of the scintillators 
%        ($t_1^A$, $t_1^B$ and $t_2^A$, $t_2^B$ for the first and second %strip, respectively).
%        The time-of-flight ($t_1 - t_2$) determines $\Delta LOR$, the %distance of the
%        annihilation point from the center of the line of response (LOR).
%    \label{fig:two_strips}
%    }
%\end{figure}

%The principle of operation is explained pictorially in %Fig.~\ref{fig:two_strips} showing two plastic strips.
%Simulated and registered spectra are shown in Figure~\ref{fig::spectra}.
%Time values ($t^A$, $t^B$) of scintillation signals arrivals at the two ends allow
%to reconstruct the photon's hit time ($t_{hit}$) and displacement from the strip center~($\Delta z_{hit}$):
%\begin{equation}
%    t_{hit}  = \frac{t^A+t^B}{2},\\
%    \Delta z_{hit}  = \frac{(t^A -t^B)\cdot v}{2},
%\end{equation}
%where $v$ denotes the effective light signal velocity in the plastic %scintillator,
%which e.g. for the strip with the cross section of 5 mm $\times$ 19 mm $\times$
%300~mm  is equal to $\sim$12.6~cm/ns~\citep{Moskal:2014sra}. 

The uncertainty of the hit-time determination and hence the Time-Of-Flight (TOF) resolution 
depends on the length  of the scintillator strip and the Coincidence Resolving Time (CRT) varies from  CRT $\approx$ 0.230~ns to CRT $\approx$ 0.440~ns
when the length changes from 15~cm to 100~cm~\citep{Moskal:2016ztv}, and it may be improved
%However, there is still room for improvement of CRT 
by using a matrix of silicon photomultiplier (SiPM) readout.
In Reference \cite{Moskal:2016ztv} it was shown that e.g. for the $2\times 5$ SiPM  matrix 
at two ends of the scintillator strip the coincidence resolving time changes from 
CRT $\approx$ 0.170~ns to CRT $\approx$ 0.365~ns   when  extending  an axial field-of-view  
(AFOV)  from  15~cm  to  100~cm~\citep{Moskal:2016ztv}.
%There is still further room for a substantial improvement of the time resolution by application 
%of other analysis methods as e.g. compressive sensing theory or Tikhonov regularization methods~\citep{Raczynski:2014poa,Raczynski:2015zca,Raczynski:2017PMB}. 

It is worth emphasizing that even higher time resolution of CRT~=~ 0.100~ns was already achieved  with the small 
$3  \times 3  \times 5 \mbox{ mm}^3$ LaBr3:Ce(5\%) 
crystals read out with the SiPMs~\citep{Schaart:2010}. 
More recently a detector design with  SiPM has been reported,
with CRT = 0.085~ns for $2  \times 2 \times 3 \mbox{ mm}^3 $ LSO:Ce codoped 0.4\%Ca crystals and CRT of  0.140~ns for $2  \times 2  \times 20 \mbox{ mm}^3$ crystals with the length as used in the current PET tomographs~\citep{PMB_Nemallapudi}. 
%Moreover, it is claimed by~\cite{Lecoq:2017} that
%there are no physical barriers of reaching the resolution of 10~ps in the future. 
%However, hereafter 
Therefore, in section~\ref{sec::MC} we will perform feasibility 
studies simulating the possibility of positronium imaging under
conservative assumption that CRT is equal to 0.140~ns, as already
achieved for the solutions applicable in the current PET detectors.

\subsection{Event selection criteria and background reduction}

\begin{figure}[tp]
    \centering
        \includegraphics[width=0.9\textwidth]{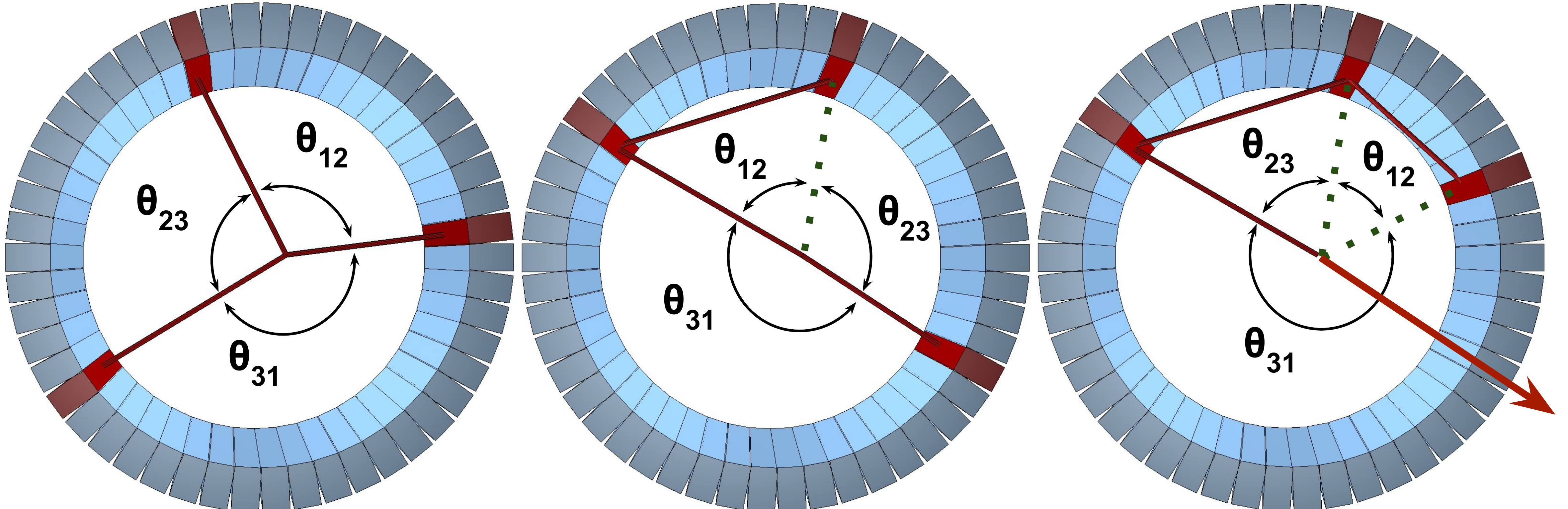} 
        \caption{
        Typical topology of \ops (left panel) and background events originating from $2\gamma$ annihilations (middle and right panels). 
       Red solid lines denote actual photons, the red arrow indicates not registered photon
       while green dotted lines  indicate reconstructed photon candidates.
       The numbering of  photons is such that $\theta_{12}< \theta_{23} <\theta_{31}$. 
        \label{fig::scatterings}
        }
\end{figure}

An event useful for reconstruction of the o-Ps image is shown pictorially in Fig.~\ref{fig::detector_with_ops}. It consists of four registered 
gamma interaction points in the scintillator strips (later on referred to as \textit{hits}),
three of them originating from the \ops annihilation and the fourth one  from the prompt gamma. 
%The simplest method of the prompt gamma identification is based on the energy loss criterion.
% From Figure~\ref{fig::spectra} it can be inferred that  when choosing a hit with the energy loss $E_{dep} > 370$~keV, the prompt gamma can be identified with 66\% and 87\% selection efficiency for $^{44}Sc$ and $^{14}O$, respectively. 
After identification of the prompt gamma (e.g. using energy loss criterion $E_{dep} > 370$~keV), the three remaining hits with $E_{dep} < 370$~keV become then candidates of photons originating from the \ops decays useful for the reconstruction of annihilation position and time  as indicated in Fig.~\ref{fig:gps_reconstruction}. However, due to the secondary Compton scattering in  plastic scintillator (discussed in more detail in references~\citep{Kowalski2015,Kowalski2016,Kowalski2018,Kaminska:2016fsn}),
the  $2\gamma$ events may mimic  registration of $3\gamma$ ones, as indicated pictorially in Figure~\ref{fig::scatterings}.
Yet, we can disentangle true and false $3\gamma$ events  based on the  (i) relation between  hit positions and hit times 
and (ii) angular correlation of relative angles between the direction of annihilation photons. 

%Signal and background events are presented in  Fig.~\ref{fig::scatterings}.

For example, events for which both photons from 
$e^+e^- \to 2\gamma$ annihilation are registered and one of them is scattered causing signals in two detectors (middle panel of Fig.~\ref{fig::scatterings}) satisfy the condition  
$\theta_{12} + \theta_{23} = 180^\circ$ 
and are concentrated along the anti-diagonal of the $\theta_{12}$ vs. $\theta_{23}$ plot. 
%Events where one   photon from $e^+e^- \to 2\gamma$ annihilation is missing detection, and another one undergoes two scatterings, are mostly localized below the anti-diagonal line. Contrary to that, the actual \ops events (signal) are localized above the anti-diagonal since 
%$\theta_{12} + \theta_{23} > 180^\circ$ 
%due to the momentum conservation. 
%The background due to the Compton scattering in the detectors 
%can be further reduced significantly comparing the time differences 
%between the hits with the time of flight needed by light to pass
%the distance between the hits. Assuming time resolution of  CRT~=~0.140~ns, 
%the secondary hits can be defined by requiring their distance from the primary hit to exceed~4~cm. 
%Detailed discussion of the background reduction power of these 
%criteria is beyond the scope of this article. Such 
Detailed studies were done in Ref.~\cite{Kaminska:2016fsn} and show that the application of the angular correlation criterion suppresses the background from $2\gamma$ by the factor of about 10$^{4}$ while rejecting only 3\% of true $3\gamma$ signal events and that further requirements based on the relation between hit position and hit time may reduce the background by another few orders of magnitude.
 
Therefore, even though the $3\gamma$ events are expected to constitute only about 0.5~\% of $2\gamma$ events,  the background due to the $2\gamma$ annihilation associated with the secondary scattering in the detector can be reduced to a negligible level. 
Hence, in the next section we present simulations assuming that the hits from the \ops annihilation can be identified and we will focus on the possibilities of  the positronium lifetime image reconstruction, assuming that the time resolution of the detector amounts to CRT~=~0.140~ns.

\subsection{Monte Carlo simulations}
\label{sec::MC}

The feasibility studies of positronium imaging are performed based on the virtual 
cylindrical phantom placed inside the J-PET tomograph as shown in Fig.~\ref{fig:oPs_detector}. 
A cylindrically shaped phantom with radii of 
$R_{inner}=10.0$~cm, $R_{outer}=10.1$~cm and $z=10.0$~cm is
placed along the $z$ axis of the detector (see Figure~\ref{fig:oPs_detector}, right panel). 
The volume of the phantom was divided into 2 parts, each corresponding to the tissues 
characterized with  different average lifetimes of ortho-positronium  atoms.
The values of average lifetimes  of 2200~ps to 2300~ps were chosen. These values are within the range of lifetimes 
expected for the ortho-positronium in the human body (as discussed in sections 
\ref{sec::introduction} and \ref{sec::positronium_in_body}). 
%The exact values of used mean lifetimes are listed
%in Table~\ref{tab::lifetime} and are indicated in Figure~\ref{fig::cylinder:generated}.

%\begin{table}[b]
%    \caption{Values of simulated o-Ps lifetime as a function of azimuthal %coordinate along the simulated cylinder. }
%    \label{tab::lifetime}
%    \resizebox{\textwidth}{!}{
%        \begin{tabular}{|c||c|c|c|c|c|c|c|} \hline
%            $\alpha \in$ & $(0,\frac{2\pi}{7})$ & %$(\frac{2\pi}{7},2\cdot\frac{2\pi}{7})$ & $(2\cdot %\frac{2\pi}{7},3\cdot\frac{2\pi}{7})$
%            & $(3\cdot \frac{2\pi}{7},4\cdot\frac{2\pi}{7})$ & $(4\cdot \frac{2\pi}{7},5\cdot\frac{2\pi}{7})$
%            & $(5\cdot \frac{2\pi}{7},6\cdot\frac{2\pi}{7})$ & $(6\cdot \frac{2\pi}{7},7\cdot\frac{2\pi}{7})$ \\ \hline
%            o-Ps lifetime & 2000~ps & 2010~ps & 2030~ps & 2080~ps & 2180~ps & 2380~ps & 2880~ps  \\ \hline
%        \end{tabular}
%    }
%\end{table}

For the studies presented in this article about $3\times 10^{9}$ events 
of o-Ps decays into three photons were generated and, subsequently, the  response of the J-PET tomograph was simulated and then the image of the mean lifetime of positronium was reconstructed. 
The simulations were conducted in the following steps: 
For each event (i) the position of the annihilation was generated, assuming a homogeneous density distribution within the cylindrical phantom; 
(ii)~the lifetime of a given o-Ps atom was generated with the exponential probability density distribution with the mean lifetime depending on the position in the phantom;
%as indicated in Table~\ref{tab::lifetime}; 
(iii) the momentum vectors of photons from the \ops decay were generated 
according to the predictions of the 
Quantum Electrodynamics~\footnote{The positronium has a small boost along the initial positron momentum. Due to this the annihilation photons' momenta slightly deviate
from co-planarity in the detector frame of reference. 
This effect is included in the simulations and its contribution is negligible
with respect to the resolution achieved so far~\citep{Gajos:2016nfg}.}; 
(iv) hit-position and energy deposition of each photon in the plastic scintillators of the J-PET tomograph were simulated taking into account the cross sections for the Compton interactions of gamma photons in the plastic scintillators;
(v) the experimental resolutions for the registration of time and position of gamma photons were accounted for by smearing
the values of generated hit-times and hit-positions with the resolution
functions of J-PET, assuming the CRT value of 0.140 ns; (vi) for each {\it registered} event, a time and position of o-Ps annihilation was reconstructed using the trilateration 
method~\citep{Gajos:2016nfg} and the image of mean o-Ps lifetime was created.  
The simulation methods for the above listed steps from (i) to (v) were described in details in the previous publications~\citep{Moskal:2016ztv,Kaminska:2016fsn,Moskal:2014sra,Moskal:2014rja}.
%However, for completeness we will briefly report  them hereafter and  the next section will be devoted to step (vi).

\begin{figure}[htp]
    \centering 
    	\includegraphics[width=0.35\textwidth]{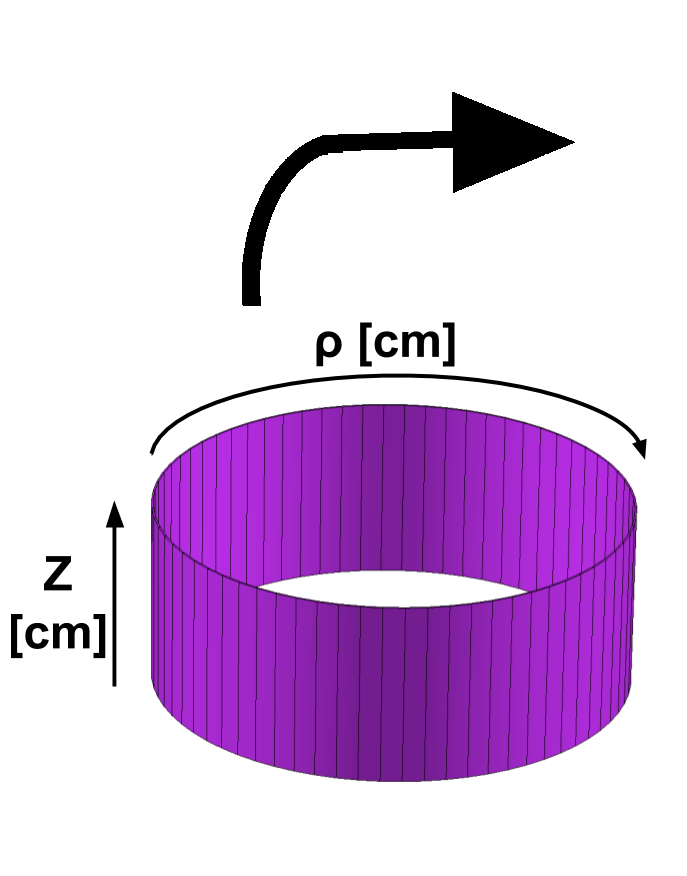}
        \includegraphics[width=0.49\textwidth]{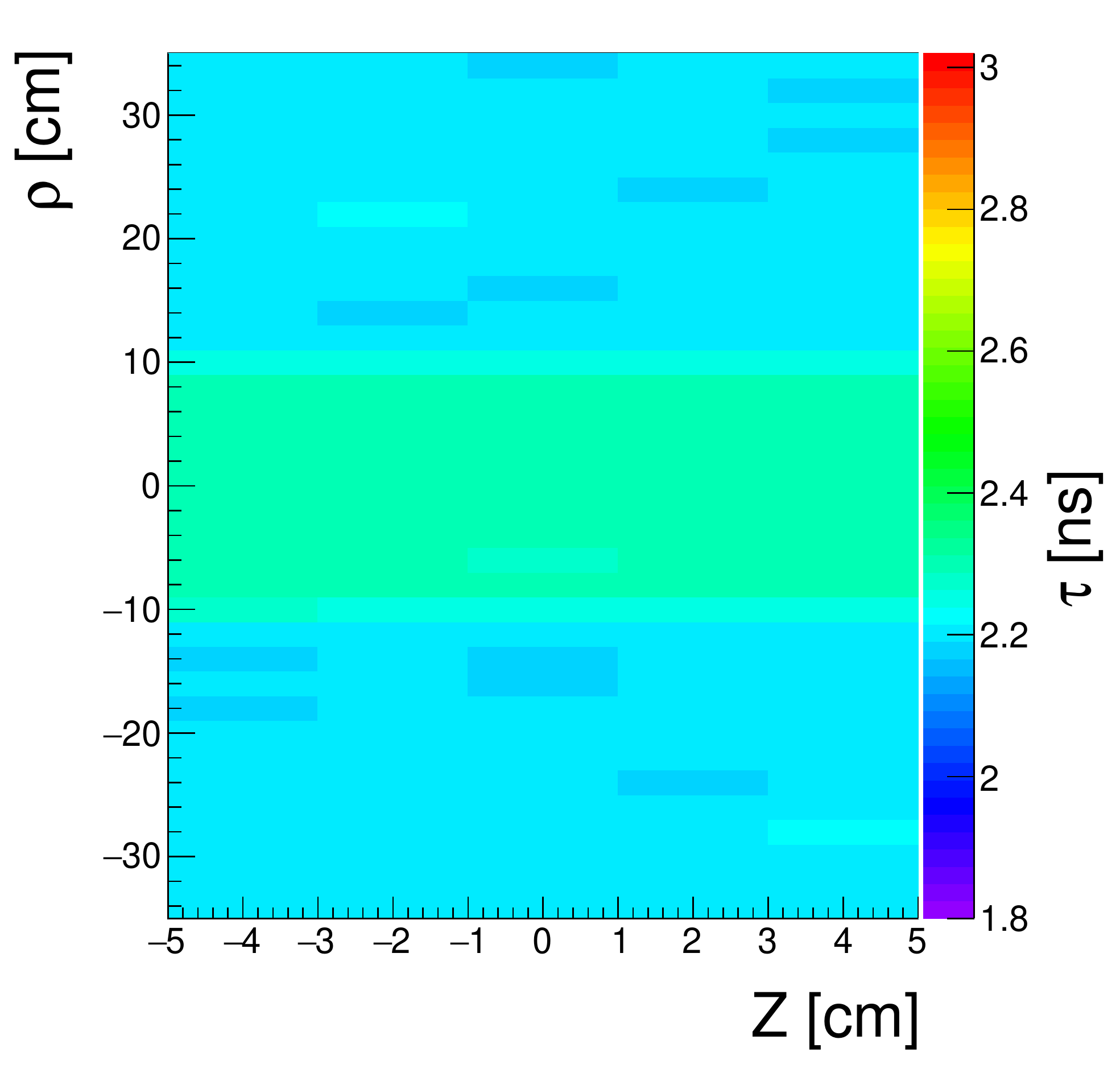}
        \caption{Generated distribution of mean ortho-positronium lifetime
        in the phantom as a function of $z$ and $\rho$ coordinates.
        Pixel size is 2~cm $\times$ 2~cm.
        Radiopharmaceutical is uniformly distributed,
        however the phantom is composed of two different materials in which
        the lifetime of ortho-positronium atoms is equal to 2300~ps (green area) and 2200~ps (blue area). 
        \label{fig::cylinder:generated}
        }
\end{figure}

% (i) i (2)
 At the step of generation of the position of annihilation and the decay time of o-Ps in the phantom, the area of the phantom was divided into pixels with dimensions of $2 \mbox{ cm} \times 2 \mbox{ cm}$. The generated mean lifetime distribution as a function of the $z-\rho$ coordinates along
the cylinder is shown in Figure~\ref{fig::cylinder:generated}. 
A colour code indicates the mean 
life-time of o-Ps calculated in each pixel, using lifetime values of ortho-positronium atoms generated in the pixel. The differences in the mean lifetimes observed within
each of two sections  are due to the finite statistics of generated events.
Thus, Figure~\ref{fig::cylinder:generated} presents an image of o-Ps mean lifetimes which could be reconstructed in the case of an ideal detector, assuming that all of the annihilations events were registered with an ideal resolution.

In addition, we preformed a simulation positioning six point annihilation sources  in the arrangement required by the  National Electrical Manufacturers Association~\citep{NEMA:2012} norm: a six sources (cylinders $r=0.5$~mm, $z=1$~mm) were placed as presented in Table~\ref{tab:nema}. Simulations included response of the detector using time and position resolutions as discussed in the text of the article.
\begin{table}[h!]
\centering
\caption{Coordinates of simulated "point-like" sources positioned according to the NEMA norm. Each source is characterized by a different o-Ps lifetime.}
\label{tab:nema}
\begin{tabular}{|C{0.2\textwidth}|C{0.4\textwidth}|C{0.2\textwidth}|} \hline
 Position & Coordinates [cm] & Simulated o-Ps lifetime [ns]\\ \hline
  1 & (1,0, 0 )     & 2.0  \\  
  2 & (10,0, 0)     & 2.4 \\
  3 & (20,0, 0 )    & 2.8 \\
  4 & (1,0, 18.75)  & 2.2 \\
  5 & (10,0, 18.75) & 2.6 \\
  6 & (20,0, 18.75) & 3.0 \\ \hline
\end{tabular}
\end{table}

\section{Results}
\label{sec::results}

\begin{figure}[!htbp]
    \centering
        \includegraphics[width=0.49\textwidth]{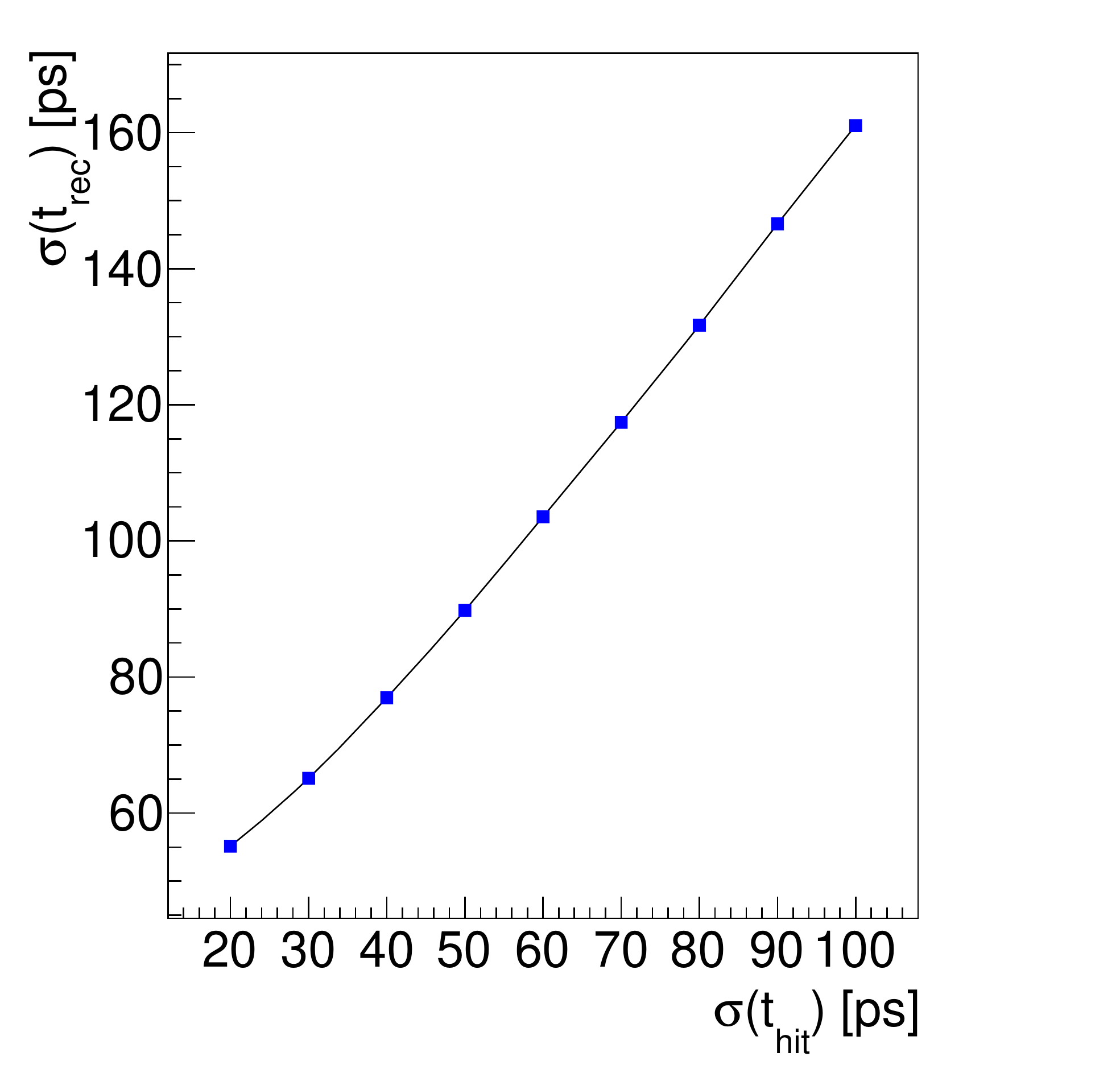}
        \includegraphics[width=0.49\textwidth]{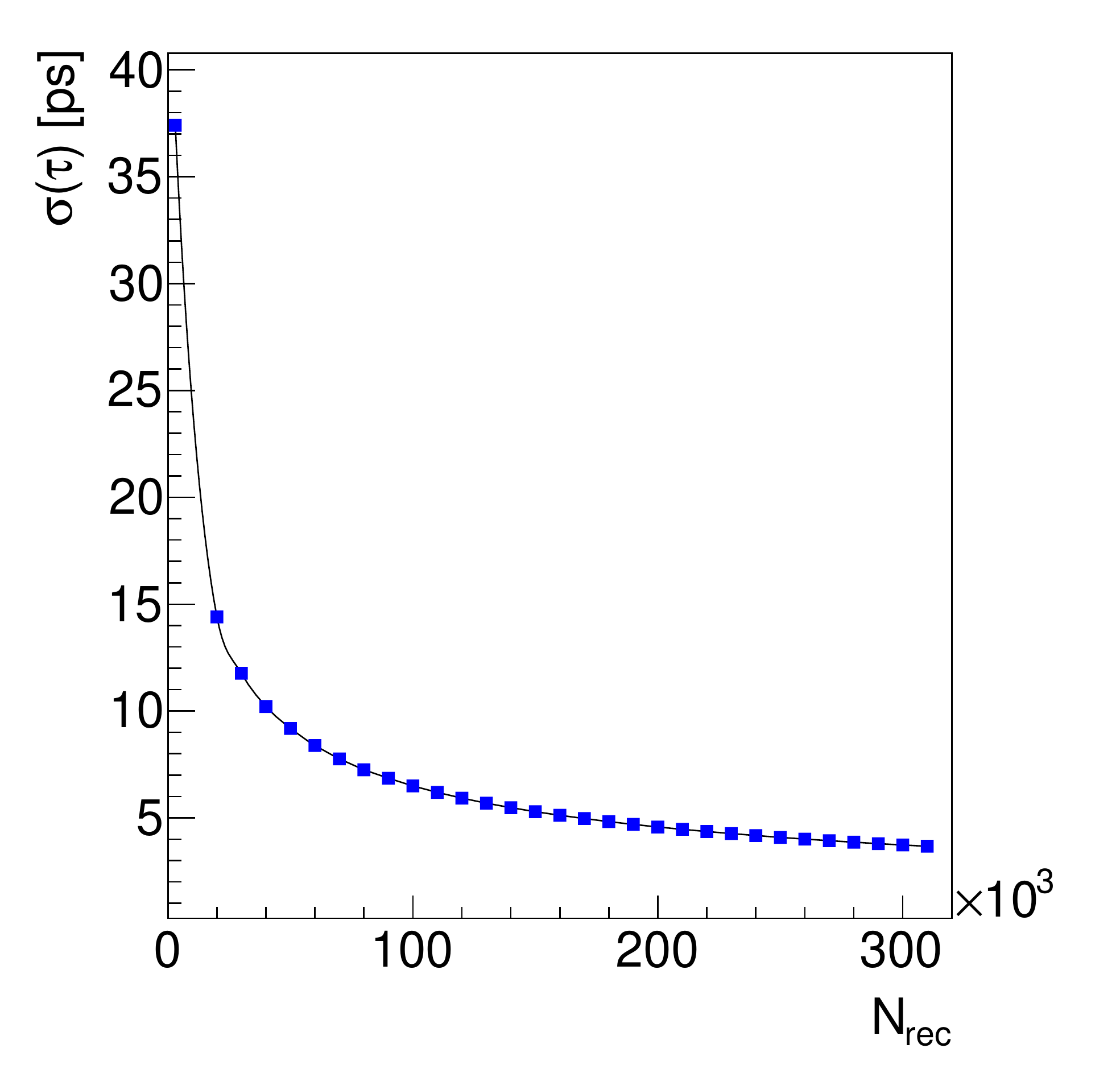}
        \caption{ 
        Left: Resolution of ortho-positronium reconstructed decay time obtained by the trilateration method as a function of detector time resolution. Achievable resolution strongly depends on detector timing capabilities. 
        Right: Resolution of the mean lifetime determination as a function of detected entries in a single voxel. An example for $\sigma(t_{hit})$~=~40ps (corresponding to CRT~$\approx$~140~ps) and $\tau$~=~2010~ps  is presented.     
        \label{fig::figure11}
        }
\end{figure}

\begin{figure}[!htbp]
    \centering
     \includegraphics[width=0.49\textwidth]{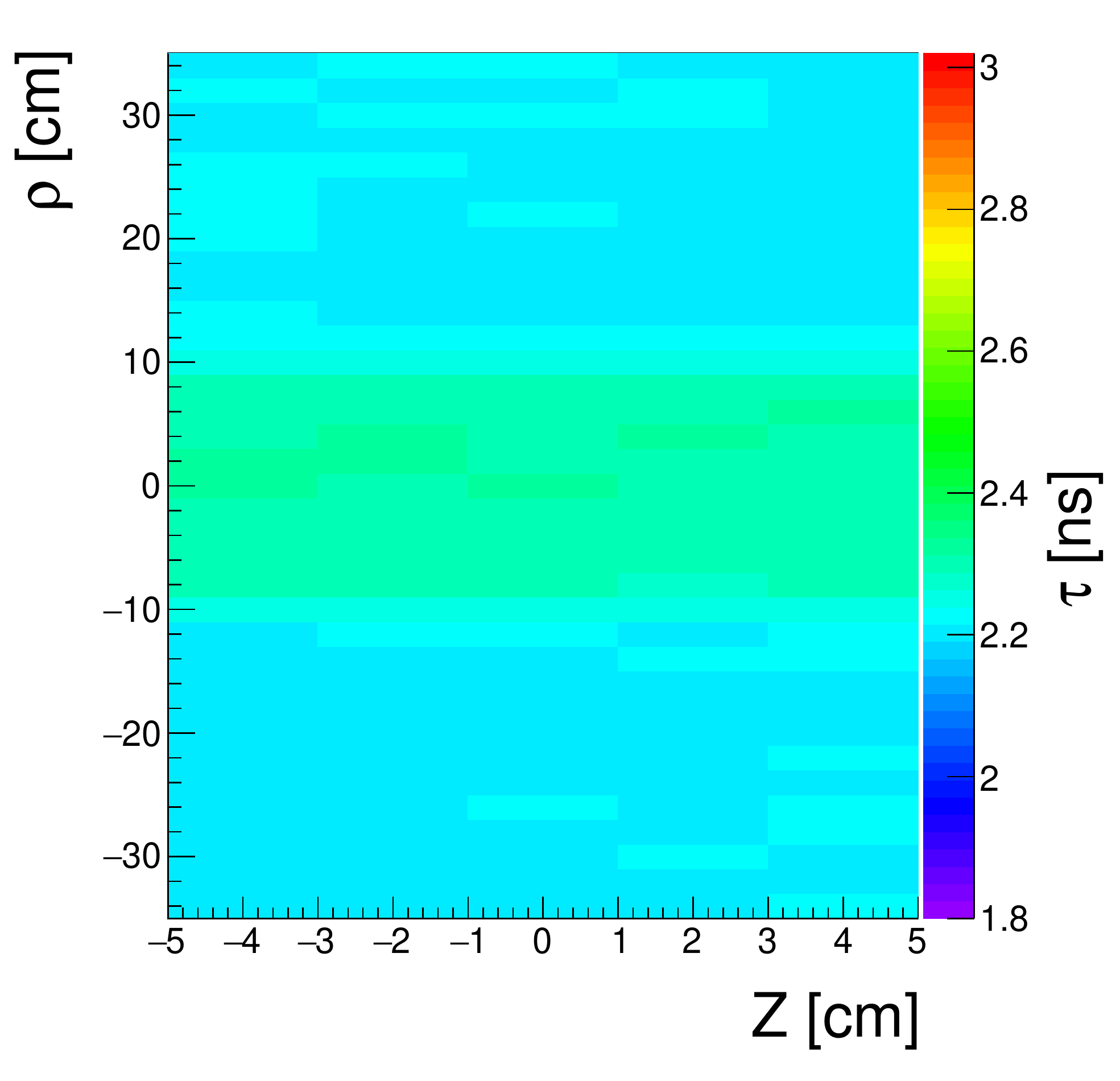}
     \includegraphics[width=0.49\textwidth]{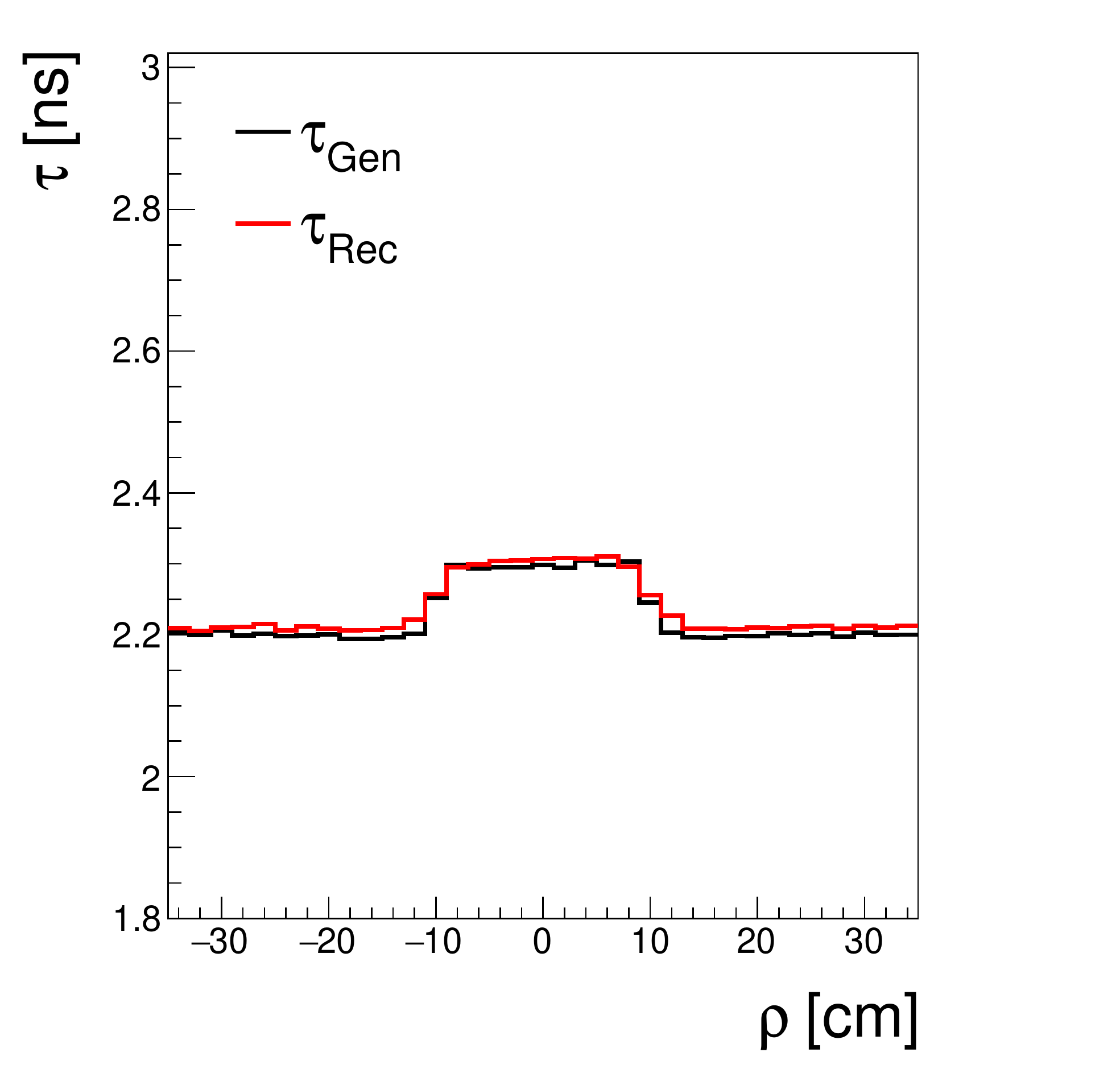} 
        \caption{
            Left:
            Reconstructed image of the phantom composed of two materials 
            in which the lifetime of ortho-positronium atoms is equal to 2200~ps (blue area) and 2300~ps (green area).
            %(see Table~\ref{tab::lifetime}).
            Generated image is shown in~Figure~\ref{fig::cylinder:generated}.
                The o-Ps annihilation time is  reconstructed by using the trilateration method and
        assuming time resolution of $\sigma(t_{hit})$~=~40~ps corresponding to CRT of about 140~ps,
        as discussed in the introduction.
       Right: Comparison between generated (black line) and reconstructed
       (red line) o-Ps lifetime image
       shown as a function of transverse radius.
         }
 \label{fig::rec_cylinder}
\end{figure}

\begin{figure}[!htbp]
    \centering
    \begin{subfigure}{.49\textwidth}
    \includegraphics[width=\textwidth]{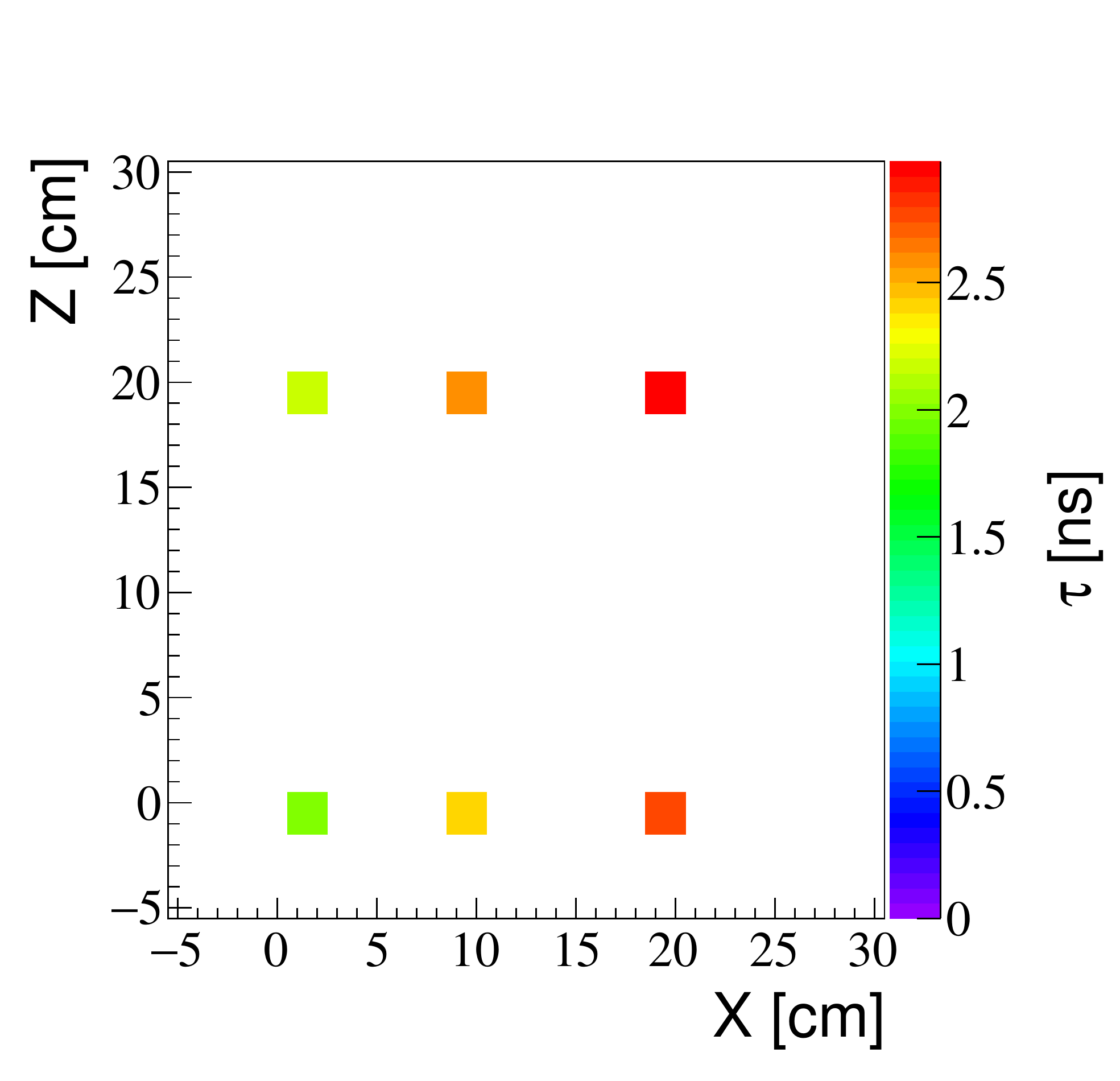}
    \caption{Generated ortho-positronium lifetime distribution.}
    \label{fig::nema_gen_lifetime}
    \end{subfigure}
    \begin{subfigure}{.49\textwidth}
    \includegraphics[width=\textwidth]{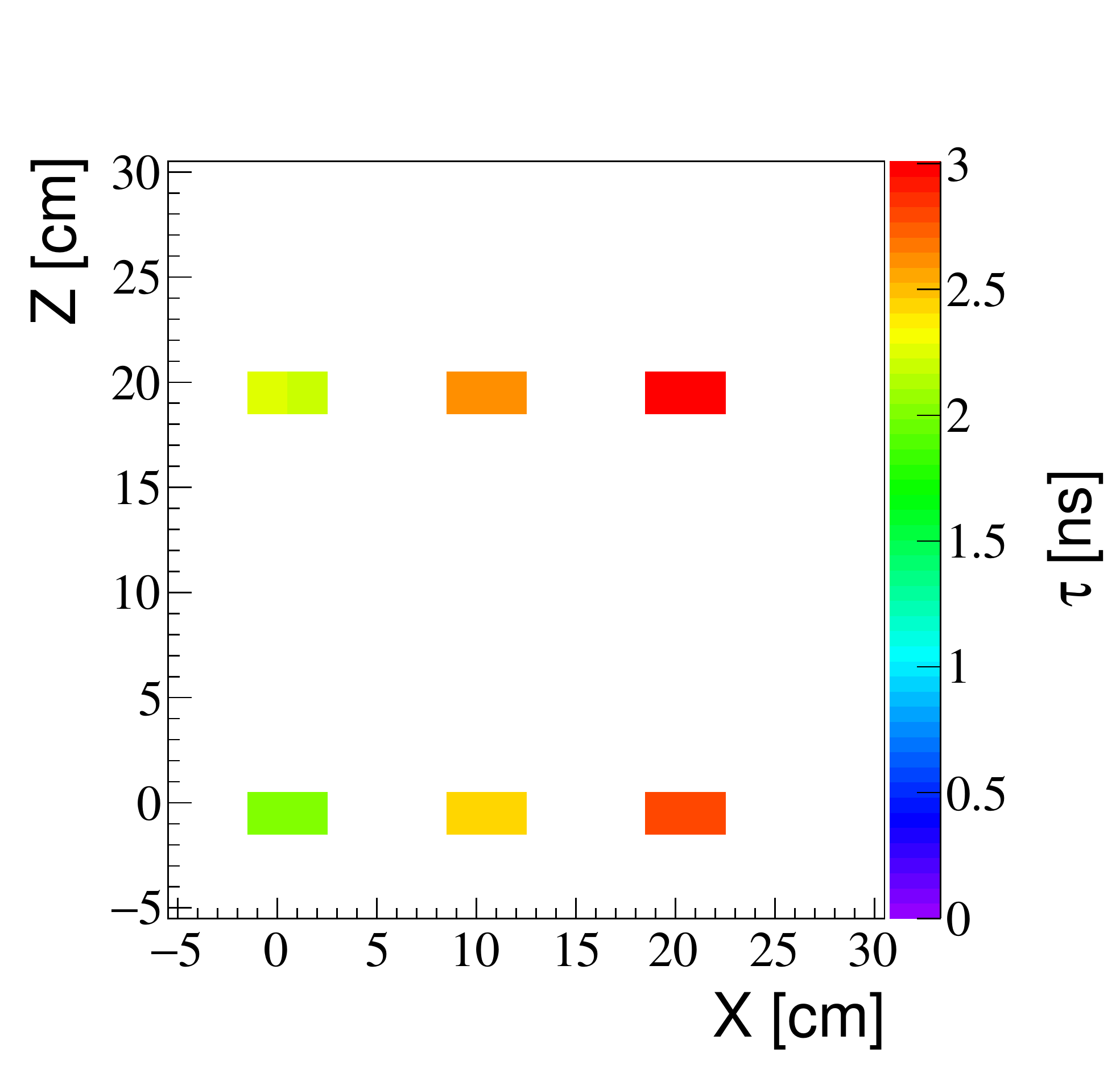}
    \caption{Reconstructed ortho-positronium lifetime distribution.}
    \label{fig:3gammaRecTau}
    \end{subfigure}
    \begin{subfigure}{.49\textwidth}
    \includegraphics[width=\textwidth]{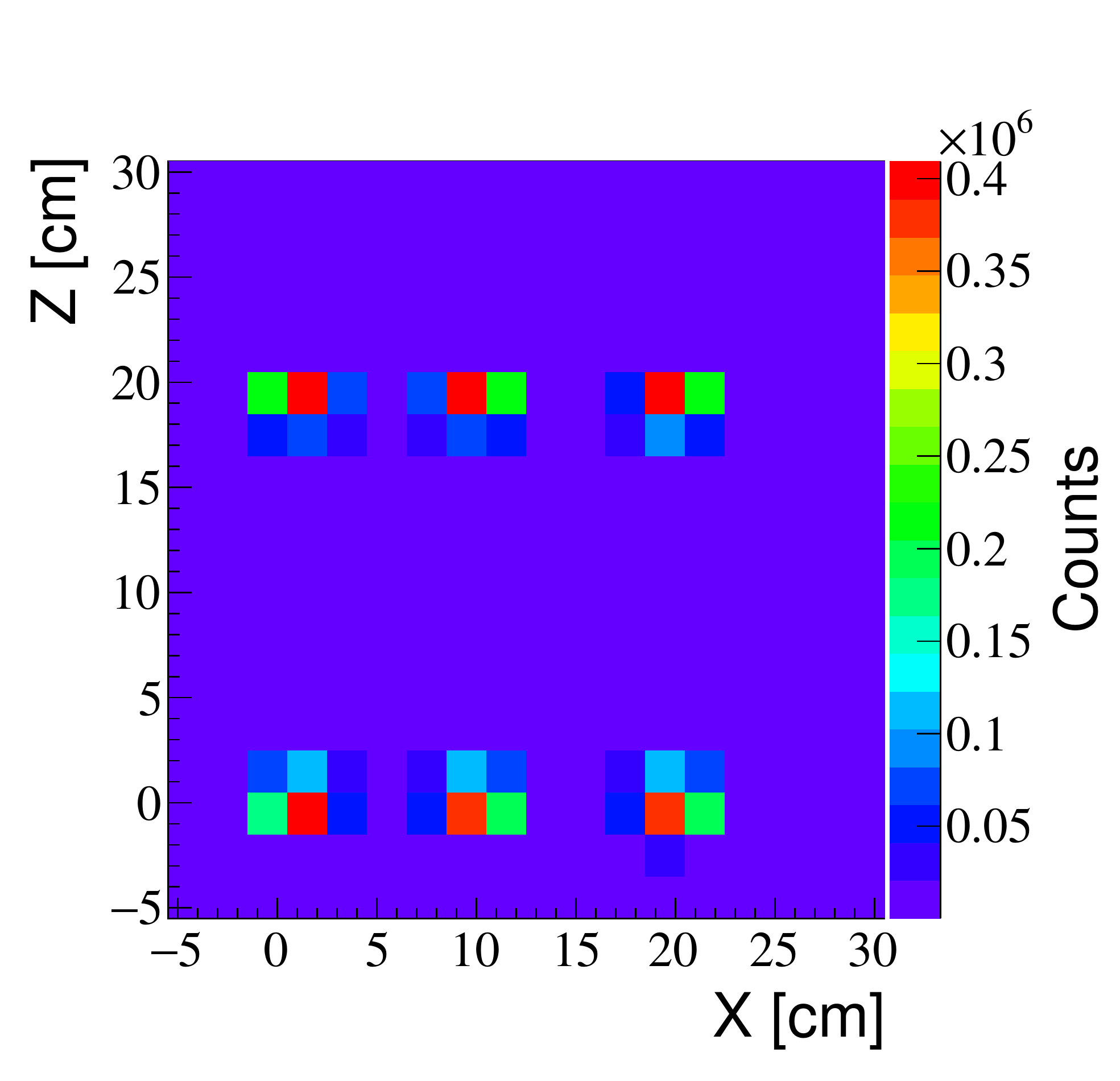}
    \caption{Reconstructed distribution of annihilation point spatial coordinates.}
    \label{fig::nema_vtx}
    \end{subfigure}
        \begin{subfigure}{.49\textwidth}
        ~\vspace{50pt}\\
    \includegraphics[width=\textwidth,height=6cm]{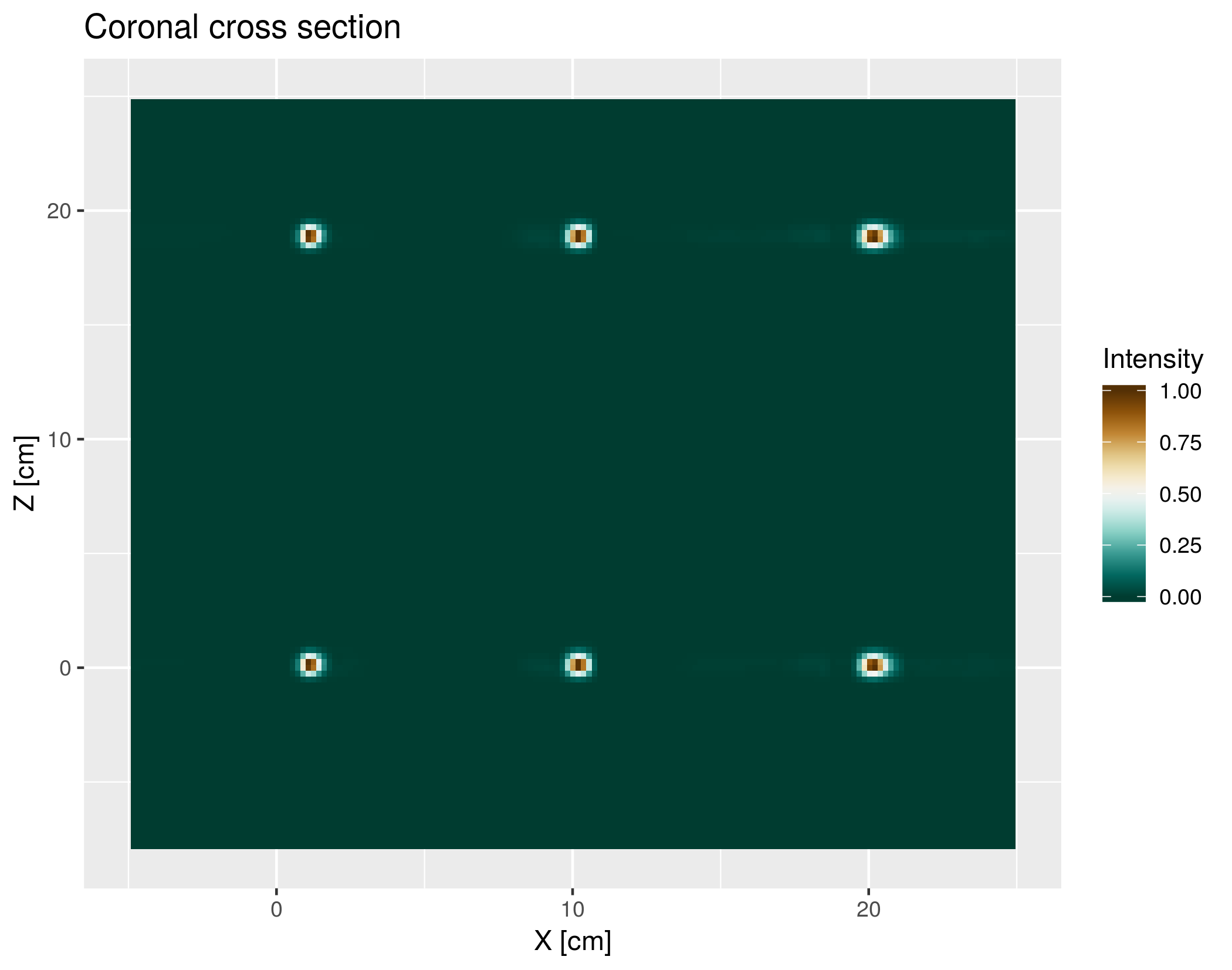}
    \caption{Reconstructed image of six sources obtained while applying the Filter Back Projection algorithm for 2$\gamma$ events.}
    \label{fig::FBP}
    \end{subfigure}
    \caption{A six point sources were placed according to the positions
    denoted in Table~\ref{tab:nema}.
    Each source has a different 
    ortho-positronium lifetime~(\subref{fig::nema_gen_lifetime}). 
    Application of $3\gamma$ trilatelation-based algorithm~\citep{Gajos:2016nfg}
    allows~to
    reconstruct distribution of spatial coordinates of annihilation points~(\subref{fig::nema_vtx}) 
    as well as
    the mean lifetime distribution of the ortho-positronium~(\subref{fig:3gammaRecTau}). 
    The voxel size is equal to $2~cm~\times~2~cm~\times~2~cm$. 
    For comparison result of standard PET imaging  
    for $2\gamma$ events is shown~(\subref{fig::FBP}).
      }
    \label{fig:3gammaRec}
\end{figure}

As a result of simulated measurements and data selection described in the previous sections
for each registered \ops event a hit-time and hit-position of
the annihilation and deexcitation photons are determined. 
For each event these enable a reconstruction of time  and position of the annihilation point 
using the trilateration method~\citep{Gajos:2016nfg}, as well as the lifetime of ortho-positronium  estimated as a difference between the emission time of annihilation photons and emission time of the prompt gamma.
For the purpose of positronium 
mean-lifetime image reconstruction
an object is divided into voxels in which \ops annihilations are recorded. 
In each voxel 
the mean lifetime is  estimated by the arithmetic mean of individual lifetimes.
Reconstruction time-resolution of a single ortho-positronium lifetime strongly depends 
on detector timing capabilities (see left panel of 
Figure~\ref{fig::figure11}) while the resolution of mean 
lifetime determination depends additionally on the number of detected events 
(see right panel of Figure~\ref{fig::figure11}). 
The  mean-lifetime resolution of about 
40~ps is achievable for about 3000 events per voxel. 

The left panel of  Figure~\ref{fig::rec_cylinder} shows the image of positronium lifetime reconstructed, assuming 
the CRT of 140~ps (corresponding to $\sigma(t_{hit})$ of about 40~ps).   
The original (ideal) image is shown  in Figure~\ref{fig::cylinder:generated}. 
The reconstructed mean lifetime image (Figure~\ref{fig::rec_cylinder})
indicates, that for CRT of about 140~ps the method allows to distinguish between
o-Ps lifetime differing by 100~ps.
Right panel of Figure~\ref{fig::rec_cylinder}
compares projections of the generated and simulated images.
It indicates that the reconstructed positronium lifetime image is in a very good agreement with the generated one.
The achieved time resolution for positronium mean
lifetime includes also uncertainties due to the depth
of interaction (which can be estimated to about 70~ps 
per event~\citep{Moskal:2016ztv}. There is still a room for the future improvement of both the 
time resolution and accuracy of DOI determination (see e.g. \citep{Lecoq:2017,vanDam2011,pizzichemi}.
%Figure
%\label{fig::lifetimedetermination} shows that for 3000 events reconstructed per voxel a mean lifetime resolution 
%of about 40~ps ($\sigma$) is expected.

Results of the test of the discussed image reconstruction method on the "point-like" sources arranged according to the NEMA norm is presented in Figure~\ref{fig:3gammaRec}. In Figure~\ref{fig:3gammaRec} the  coronal ($XZ$) cross-section along $y=0$~cm is shown, for all six positions of the source. 
In each source a different o-Ps lifetime were assumed (see~Table~\ref{tab:nema}). 
Reconstructed and generated ortho-positronium lifetime are
in agreement (compare Figure \ref{fig::nema_gen_lifetime} and \ref{fig:3gammaRecTau}).
Spread originates from uncertainty of the reconstructed 
spatial coordinates of the annihilation point. 

The conventional PET imaging  on simulated $2\gamma$ events (see~Figure~\ref{fig::FBP})
and trilateration-based reconstruction
of ortho-positronium decays into $3\gamma$ photons
(see~Figure~\ref{fig::nema_vtx})
were preformed. 
The Filtered Back Projection (FBP) is a standard approved by NEMA for the estimation of spatial resolution. Therefore, we employed 3D Reprojection (3DRP) FBP algorithm for image reconstruction, implemented in STIR framework~\citep{stir}.
STIR does not support multilayer geometry, hence all events were remapped onto a single layer of radius $R=43.73$~mm, comprising the same $7$-mm wide strips, $384$ in total (an ideal scanner, defined in previous works~\citep{shopa2017}.
After the smearing of times and $Z$-positions of hits were applied, a SAFIR package was utilised~\citep{BECKER2017} 
for data conversion from list-mode into a STIR-compatible format. 
In Figure~\ref{fig::FBP}, the coronal ($XZ$) cross-section along $y=0$~cm is shown, for all six positions of the source, aggregated into one image. It is important to note that transverse spatial resolution is expected to be higher for the initial 4-layer scanner, since it provides better sinogram sampling in projection space.

%\section{Discussion}
%\label{sec::discussion}

\begin{figure}
\centering
 \includegraphics[width=0.49\textwidth]{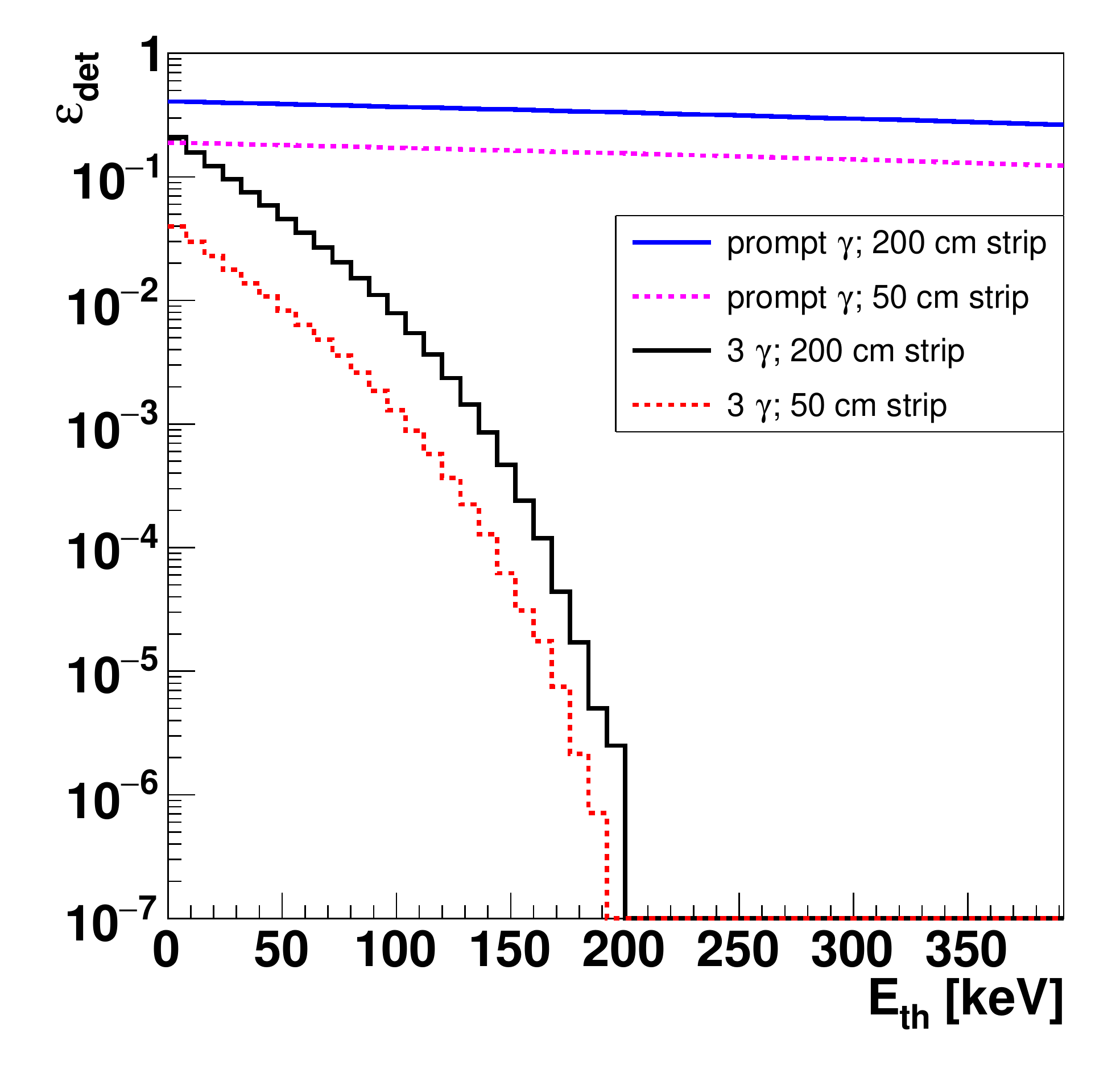}
 \caption{Registration  efficiency for prompt gamma (upper lines) and \ops as a function of the energy-loss threshold
       (determined  at the center of the tomograph, taking  into
        account geometrical acceptance, 
        probability of gamma photons registration in 
        the plastic scintillator and J-PET detector resolution).
        The dashed and solid lines
        indicate efficiencies for 50~cm and 200~cm 
        strip length, respectively.}
  \label{fig::efficiency}
 \end{figure}

The efficiencies for the registration of three photons from the \ops annihilation
with the four-layer J-PET with  AFOV of 50~cm and 200~cm are shown in the  Fig.~\ref{fig::efficiency}. 
The efficiencies are plotted as a function of the energy loss threshold. For the threshold of 25~keV the determined efficiency for \ops amounts to  2\% and  9\%  for AFOV of 50~cm and 200~cm, respectively.
Thus, about $2 \times 10^8$ $3\gamma$ events can be collected assuming that: (i)~activity of 370~MBq (10 mCi) is administered to the patient~\citep{Cherry2018}, (ii)~the data is acquired during 20 minutes according to the standard total body protocol, (iii)~the sensitivity for $3\gamma$ detection with total-body J-PET amounts to 9\%, 
(iv)~the fraction of 3$\gamma$ annihilations is equal to 0.5\%.

In addition, taking into account the sensitivity for the detection and selection (energy loss larger than 370~keV) of the prompt gamma equal
to about 27\% (blue solid line in the right panel of Figure~\ref{fig::efficiency}), one obtains $5 \times 10^7$ registered $3\gamma$ events
in coincidence with the prompt gamma. 
This gives a promising perspective, especially in view of the recent works showing that for the current PET with AFOV~=~25.5~cm  (Siemens mMR PET/MR scanner)
even $10^6$ counts in the scan could be sufficient to provide good images~\citep{Yan2016}. 
$5 \times 10^7$ registered events result  in, on average, about 700 events per cubic centimeter of the examined patient.
%in terms of SNR, CNR, bias and nois
%
%napisali w konkluzji ze
	%"Initial results with the given data of 16 patients diagnosed as TB demonstrated that 5 × 106 counts %in the scan could be sufficient to yield good images in terms of SNR, CNR, bias and noise."
However, the effective imaging sensitivity may be significantly enhanced by improving spatial resolution of the determination of the annihilation position for a single event. The signal-to-noise ratio (SNR) in the conventional PET image is proportional to the square root of sensitivity~S,
%activity~A and time of acquisition~T:
%$SNR~\sim \sqrt{S\times A \times T}$~\citep{Charry2018},
$SNR~\sim \sqrt{S}$,
and it may be increased effectively by improving the uncertainty in the localization of the annihilation point
$\Delta X$ (SNR~$\sim \sqrt{1/\Delta X}$)~\citep{Conti2009,Eriksson2015}, 
where $\Delta X$ is proportional to the time resolution (e.g. 
CRT~=~0.5~ns gives $\Delta X~=~7.5$~cm). For the activity distributed 
homogeneously in 40~cm diameter phantom the SNR increases 
with $\Delta X$~=~7.5~cm by a factor of about 2.3~\citep{Townsend2008},
corresponding to the effective sensitivity growth by a factor of about 5.3.
Thus, improving $\Delta X$ to 1~cm would lead to a gain in sensitivity by a factor of about 40 with respect to non-TOF PET systems. 
The three photon iterative image reconstruction methods remains to be elaborated, nevertheless,
this challenge is far beyond the scope of this article.
Thus, presently, in the case of the three photon reconstruction discussed in this article, it is impossible to precisely evaluate the gain
in effective sensitivity due to the improvement of the spatial resolution of annihilation point reconstruction. 
%Therefore, we may only speculate that achieving a spatial resolution 
%for a single event below ~1~cm can also enhance effectively
%the sensitivity by a factor of about 40. 
However, the estimations of the sensitivity of about 9\% for three
photon registration with the 200~cm long J-PET,  and the estimated spatial resolution 
for a single event of about 1~cm~\citep{Gajos:2016nfg} give  promising 
perspectives for the future application of the positronium imaging even though the fraction of three photons annihilations in human body is expected to be at the level of about 0.5~\% only.

\section{Summary and perspectives}
\label{sec::summary}

In this article we presented a cost-effective method enabling reconstruction of the density
distribution of the $e^+e^- \to 3\gamma$ annihilation points, as well as reconstruction
of the image of ortho-positronium average lifetime. The proposed method is based on the 
time of signals registered with plastic scintillators and its precision relies predominantly 
on the time resolution of the tomograph. The arrival times of signals measured at the ends of 
scintillator strips are used to reconstruct the hit-times and hit-positions of registered 
gamma photons. Next, the annihilation point is reconstructed analytically, on an event by event
basis, using  trilateration method and taking advantage of the fact that all three photons 
from the decay of ortho-positronium are lying in single plane.

We explored the possibility to apply the introduced method for the reconstruction
of images of mean lifetime of ortho-positronium which are created
in patients body during the routing PET investigations. In the patient about 40\% of $e^+e^-$ annihilations proceed via formation of positronium atoms with about 0.5\%  decaying into $3\gamma$~\citep{Harpen2004,JasinskaActaB:2017}.
Though the relative rate of $3\gamma$ annihilations is low, we have argued that the positronium imaging may become feasible with the advent of the total-body PET scanners.  
 The average lifetime of positronium, as well as the fraction of its annihilations into three-photons, depend on the size of the free volumes between atoms
and there are experimental indications~\citep{JasinskaActaB:2017,JasinskaActaA:2017,Pietrzak,Liu2007} 
that it is correlated with the metabolic disorders of the human tissues. Therefore, imaging of the properties of positronium inside the body may deliver new in-vivo diagnostic information complementary to the standardized uptake value (SUV index). 

In this article we focused our attention on the feasibility studies of imaging of 
the ortho-positronium mean lifetime which requires application of 
radiopharmaceuticals labeled with isotopes emitting a prompt gamma (e.g. $^{44}$Sc). 
The prompt gamma is used to determine the  creation time of positronium and 
thus the lifetime may be determined event-by-event based on the reconstructed 
time difference between annihilation and creation.  
However, the introduced method may be also used to determine images of the ratio 
of the rate of $3\gamma$ to $2 \gamma$ annihilations which is also sensitive 
to the size and concentration of free voids between atoms~\citep{Jasinska-Moskal2017}.
In this case the prompt gamma is not required  and such an image may be created for all radiopharmaceuticals, independently of the kind of the used $\beta^+$ tracer.

\section*{Acknowledgments}
The authors acknowledge technical and administrative support of A.~Heczko, M.~Kajetanowicz and W.~Migda\l{}. This work was supported by The Polish National Center for Research
and Development through grant INNOTECH-K1/IN1/64/159174/NCBR/12, the
Foundation for Polish Science through the MPD and TEAM/2017-4/39 programmes, the National Science Centre of Poland through grants no.\
2016/21/B/ST2/01222, 2017/25/N/NZ1/00861, the Ministry for Science and Higher Education through grants no. 6673/IA/SP/2016,
7150/E-338/SPUB/2017/1 and 7150/E-338/M/2017, 7150/E-338/M/2018 and the Austrian Science Fund FWF-P26783.

\bibliography{jpet_imaging}

\end{document}